\documentclass{aa}  
\usepackage[varg]{txfonts}
\usepackage{hyperref}

\usepackage{upgreek}
\usepackage{graphicx}   % Including figure files
\newcommand{\be}{\begin{equation}}
\newcommand{\ee}{\end{equation}}
\newcommand{\bea}{\begin{eqnarray}}
\newcommand{\eea}{\end{eqnarray}}

\newcommand{\unit}[1]{\mbox{\boldmath $\hat{#1}$}}

\newcommand{\fc}{f_{\rm c}}
\newcommand{\teff}{T_{\rm eff}}
\newcommand{\ledd}{L_{\rm Edd}}
\newcommand{\msun}{M_{\odot}}  

\begin{document}

\title{Observational appearance of rapidly rotating neutron stars}
\subtitle{X-ray bursts, cooling tail method, and radius determination}
 \titlerunning{Observational appearance of rapidly rotating neutron stars}
   
\author{Valery~F. Suleimanov\inst{1,2,3}
\and
Juri Poutanen\inst{4,3,5}
\and 
Klaus Werner\inst{1}}
\authorrunning{Valery~F. Suleimanov et al.}

\institute{Institut f\"ur Astronomie und Astrophysik, Kepler Center for Astro and
Particle Physics, Universit\"at T\"ubingen, Sand 1, 72076 T\"ubingen, Germany\\ \email{suleimanov@astro.uni-tuebingen.de}
\and Astronomy Department, Kazan (Volga region) Federal University, Kremlyovskaya str. 18, 420008 Kazan, Russia
\and Space Research Institute of the Russian Academy of Sciences, Profsoyuznaya str. 84/32, 117997 Moscow, Russia 
\and Department of Physics and Astronomy,  FI-20014 University of Turku, Finland 
\and Nordita, KTH Royal Institute of Technology and Stockholm University, Roslagstullsbacken 23, SE-10691 Stockholm, Sweden
}

%\date{Received xxx / Accepted xxx}

\abstract
{Neutron stars (NSs) in low-mass X-ray binaries rotate at frequencies high enough  to significantly deviate from sphericity ($\nu_* \sim$200--600\,Hz). 
First,  we investigate the effects of rapid rotation on the observational appearance of a NS. 
We propose analytical formulae relating gravitational mass and equatorial radius of the rapidly rotating NS to the mass $M$ and radius $R$  of a non-rotating NS of the same  baryonic mass using accurate fully relativistic computations. 
We assume that the NS surface emission is described by the Planck function with two different emission patterns: the isotropic intensity and that corresponding to the electron-scattering dominated atmosphere. 
For these two cases we compute spectra from an oblate rotating NS observed at different inclination angles using the modified oblate Schwarzschild (MOS) approximation, where light bending is computed in Schwarzschild metric, but frame dragging and quadrupole moment of a NS are approximately accounted for in the photon redshift calculations. 
In particular, we determine the solid angle at which a rotating NS is seen by a distant observer, the observed colour temperature and the blackbody normalization. 
Then, we investigate how rapid rotation affects the results of NS radius determination using the cooling tail method applied to the X-ray burst spectral evolution. 
We approximate the local spectra from the NS surface by a diluted blackbody with the luminosity-dependent dilution factor using previously computed NS atmosphere models. 
We then generalize the cooling tail method to the case of a rapidly rotating NS to obtain the most probable values of $M$ and $R$ of the corresponding non-rotating NS with the same baryonic mass.
We show that the NS radius could be overestimated by 3--3.5\,km for face-on stars of $R\approx 11$~km rotating at $\nu_* =$\,700\,Hz if the version of the cooling tail method for a non-rotating NS is used. 
We apply the method to an X-ray burst observed from the NS rotating at $\nu_* \approx$\,532\,Hz in SAX~J1810.8$-$2609.
The resulting radius of the non-rotating NS (assuming $M=1.5 M_\odot$) becomes $11.8\pm0.5$\,km if it is viewed at inclination $i=60\degr$  and $R=11.2\pm0.5$\,km for a face-on view, which are smaller by 0.6 and 1.2~km than the radius obtained using standard cooling tail method ignoring rotation. 
The corresponding equatorial radii of these rapidly rotating NSs are 12.3$\pm 0.6$\,km (for $i=60\degr$) and 11.6$\pm 0.6$\,km (for $i=0\degr$).
}

\keywords{stars: neutron  -- stars: atmospheres -- methods: numerical -- stars: individual (SAX~J1810.8$-$2609) -- X-rays: binaries -- X-rays: bursts}

\maketitle

%%%%%%%%%%%%%%%%%%%%%%%%%%%%%%%%%%%%%%%%%%%%%%%%%%

%%%%%%%%%%%%%%%%% BODY OF PAPER %%%%%%%%%%%%%%%%%%

\section{Introduction}

Thermonuclear explosions (flashes) of the freshly accreting matter on the surface of neutron stars (NSs)
in low-mass X-ray binaries (LMXBs) are observed as type I  X-ray bursts \citep{1993SSRv...62..223L,SB06}. 
The most successful  X-ray observatory for X-ray burst studies was  \textit{Rossi X-ray Timing Explorer (RXTE)}. 
The large area and the high time resolution of its Proportional Counter Array (PCA)  \citep{JMR06} allowed investigation of the X-ray burst flux variability with unprecedented accuracy \citep{Galloway08}. 
Thanks to \textit{RXTE}/PCA, it was also possible to perform accurate measurements of X-ray burst spectral evolution.  
The burst spectra, usually well fitted with a blackbody, give a detailed view of the evolution of the main parameters, such as the blackbody temperature $T_{\rm BB}$, its normalization $K$, and the blackbody flux $F_{\rm BB}$ \citep[see e.g.][]{Galloway08}. 
However, local spectra of hot NS atmospheres are not actually blackbodies, but are rather close to a diluted blackbody with the surface flux proportional to the Planck function $w  \uppi B_{E} (T_{\rm c})$.
The colour temperature of the spectrum $T_{\rm c}$ is typically higher than the effective temperature $\teff$ by the colour-correction factor $\fc=T_{\rm c}/\teff> 1$. 
As a result, the dilution factor $w \approx \fc^{-4}$ has to be smaller than unity to conserve the bolometric flux. 
The observed blackbody temperature $T_{\rm BB}=T_{\rm c}/(1+z)$ is just a gravitationally redshifted colour temperature.
The blackbody normalization $K$ of the observed burst spectrum is proportional to the dilution factor $w$ characterizing properties of radiation escaping from the NS surface.

The dilution and colour-correction factors depend on the  NS luminosity relative to the Eddington luminosity, and change most significantly near the Eddington limit. 
Therefore, the most powerful X-ray bursts with luminosities exceeding the Eddington luminosity,  so-called photospheric radius expansion (PRE) bursts, have to demonstrate a significant evolution of the blackbody normalization after the moment when the photosphere settles down at the NS surface (the touchdown point). 
Indeed, this kind of evolution has been observed in some X-ray bursts and the cooling tail method was designed  to obtain the apparent NS size and the Eddington flux from the data  \citep{SPRW11,Poutanen.etal:14}. 
They were then converted to constraints on the NS basic parameters such as mass $M$ and radius $R$. 
However, due to a divergence of the Jacobian the resulting distribution of $M$ and $R$ is biased \citep{Ozel15}. 
This problem was avoided by fitting the evolution of the blackbody normalization during the cooling tail using $M$ and $R$ as parameters \citep{Nattila.etal:16,Suleimanov.etal:17}.  
A more detailed description, the history, and applications of the method are reviewed in  \citet{Suleimanov.etal:16} and \citet{DS18}.    
Further development of the method is the direct fitting of the observed spectral sequence with the model spectra of hot NS atmospheres. 
It allowed us to obtain NS radii with unprecedented accuracy of 0.5\,km \citep{Nattila.etal:17}. 
Such a narrow range of allowed NS radii already gives interesting constraints on the equation of state (EoS) of cold dense matter  \citep[see the recent review by ][]{LP16}.

The studies quoted above assume that the NS is spherical and its emission is isotropic. 
However, the discovery by \textit{RXTE} of burst oscillations \citep{Strohmayer96} and accreting millisecond pulsars \citep{WvdK98}  imply that NSs in LMXBs are rapidly rotating at a rate in the range $\nu_*\approx$ 200--600\,Hz  \citep{Watts12,PW12,Campana18}.
Rapid rotation distorts  the NS shape and produces a potentially important systematic effect on the NS  radii determination. 
Radiation from the approaching side of the NS becomes Doppler boosted shifting the emission peak to higher energies as, for example, was extensively discussed in the studies of pulse profiles from millisecond pulsars using the  so-called Schwarzschild+Doppler (S+D) \citep{PG03,PB06} and oblate Schwarzschild (OS) approximations \citep{mors07,Bogdanov19L26}. 
Rotation also affects the value of the local gravitational acceleration and therefore the value of the flux in Eddington units and as a consequence the escaping spectrum  and the colour correction. 
 
Models for rotating NS have been constructed for a rather long time, with the current state being presented by \citet{PS17}. 
The accurate shape of a rotating NS depends on the assumed EoS, but for the astrophysical applications some approximate rotating NS models as suggested by \citet{mors07} and \citet{ALGM:14} are sufficient.
The influence of the oblate form of the rotating NSs on the  pulse profiles of millisecond pulsars was considered using accurate general relativistic ray-tracing in the curved space-time  \citep[see e.g.][]{Cadeau.etal:07, NP:18}.  
The effects of rapid rotation on the observed spectra and the line profiles were studied by \citet{Baubock.etal:13, Baubock.etal:15}.  
In particular, \citet{Baubock.etal:15} computed the angle-averaged correction to the measured NS radius as  a function of the rotation frequency assuming local isotropic blackbody emission of the same temperature.
\citet{Vincent.etal:18}, on the other hand, used the local model atmosphere spectra from \citet{Majczyna:05}, but assumed no latitudinal variation. 
We note that a modest change in the gravitational acceleration leads to large change in the colour-correction factor, especially close to the Eddington limit. 
Thus, we suggest that in order to compute realistic observed X-ray burst spectra,  latitudinal variations of the local atmosphere spectra (e.g. in terms of variations of the colour-correction factor) should be accounted for.

In this paper we suggest a new method for computing spectra from rapidly rotating NSs. 
We base our approach on a slow-rotation approximation for the NS shape and space-time metric proposed by \citet{ALGM:14}. 
We introduce the modified oblate Schwarzschild (MOS) approximation where light bending is computed in Schwarzschild metric, but for calculations of the gravitational redshift and Doppler boost we approximately account for the frame dragging and quadrupole moment of a NS. 
In Sect.\,\ref{sec:rotatingNS} we propose analytical formulae connecting the gravitational mass and equatorial radius of a rotating NS to the mass and radius of a non-rotating NS of the same baryonic mass. 
We then compute the spectra of the blackbody emitting, rapidly rotating NSs and obtain the colour-correction and dilution factors as a function of rotation rate.
We extend the direct cooling tail method to rapidly rotating NSs in Sect.\,\ref{sec:method}.
The modified method is applied  to the X-ray burst from LMXB SAX~J1810.8$-$2609  in Sect.\,\ref{sec:application}. 
We summarize the results in Sect.\,\ref{sec:summary}.     
Technical details are presented in the Appendices.

\section{Observational appearance of rapidly rotating NSs}
\label{sec:rotatingNS}

One of the goals of the current work is to develop the modified direct cooling tail method applicable to rapidly rotating NSs.  
The final result of the method would be an estimation of the equivalent radius of a non-rotating NS of a given mass using observed data on the X-ray bursts from the surface of a rapidly rotating NS. 
Two additional parameters appearing in this task are the rotational angular velocity $\Omega_*=2\uppi \nu_*$ of the NS and the inclination angle $i$ between the rotation axis and the line of sight to the observer. 
The exact model of the rapidly  rotating NS depends on the assumed EoS in its inner core, which is not well known.   
Moreover, constraining it from the observations is one of our goals. 
Fortunately, the shape of a rapidly rotating NS depends on the actual EoS only slightly, if the rotation is not very rapid, $\nu_* < 700$\,Hz.  
In this case, a slow-rotation approximation can be used, and the model of a rapidly rotating NS depends only on its gravitational mass $M$, equatorial radius $R_{\rm e}$, and the angular velocity $\Omega_*$, as discussed by \citet{mors07} and  \citet{ALGM:14}. 
The approach suggested in these works is used in this paper (see Appendix \ref{sec:app1} for details).
We describe now the effects arising due to rapid rotation.

\begin{figure}
\centering
\includegraphics[width=0.80\columnwidth]{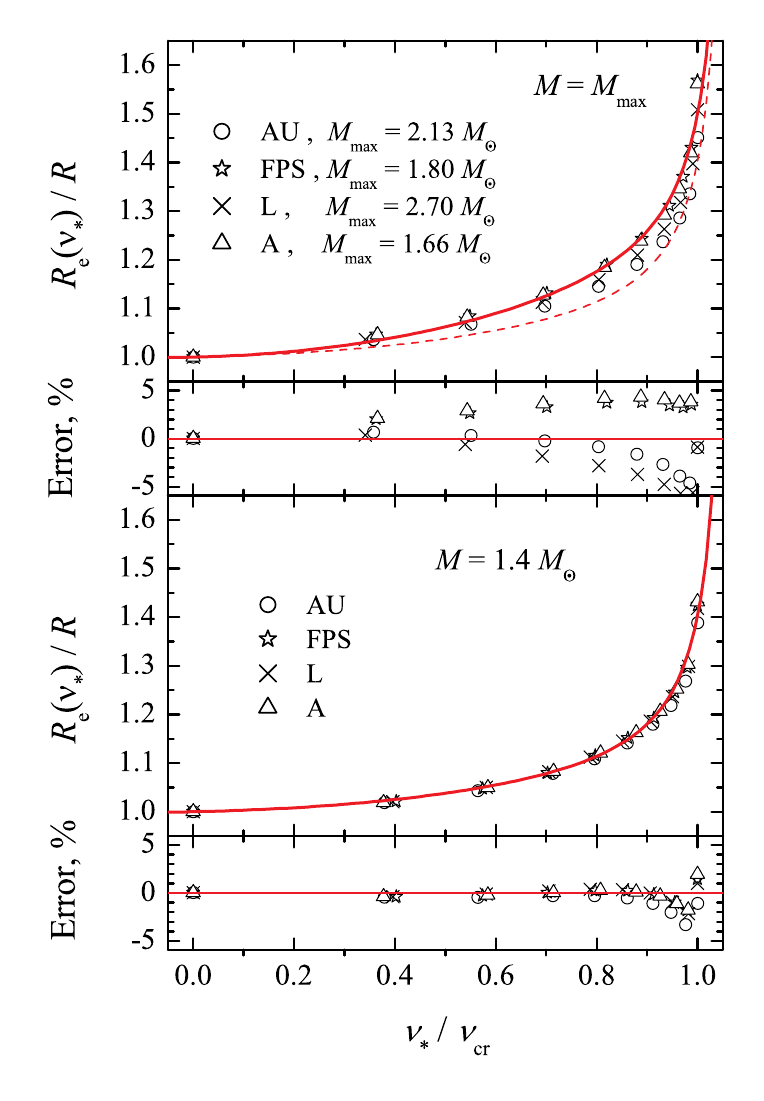}
\caption{Dependences
of the  ratio $R_{\rm e}/R$ on the relative rotational frequency for numerical NS models with various EoS \citep{CShT94} for the maximum possible mass ({\it top panel}) and the mass $M=1.4\msun$ ({\it bottom panel}). 
Red solid curves show the fits with Eq.\,(\ref{eq:RRe}) for the masses $M=2.5\msun$ ({\it top panel}) and $1.4\msun$ 
({\it bottom panel}).
The fitting curve for $M=1.4\msun$ is shown in the top panel as dashes. 
The relative errors of the fits are also shown in the narrow additional panels below the main panels.
}
\label{fig:radius}
\end{figure}

\begin{figure}
\centering{\includegraphics[width=0.80\columnwidth]{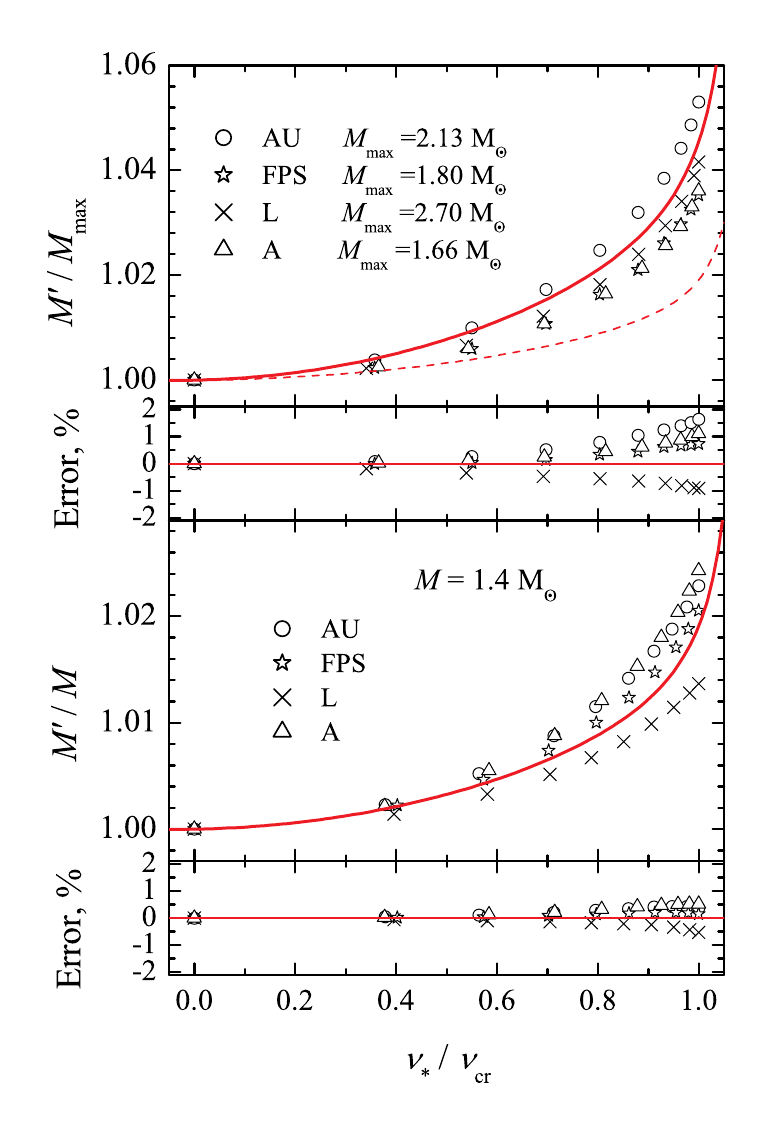}} 
\caption{Dependences of ratios $M'/M$ on the relative rotational frequency for numerical NS models with various EoS \citep{CShT94} for the maximum possible mass (top panel) and the mass $M=1.4\msun$ (bottom panel). 
Red solid curves show the fits with Eq.\,(\ref{eq:massprime}) for the masses $M=2.5\msun$ ({\it top panel}) and $1.4\msun$ 
({\it bottom panel}).
The fitting curve for $M=1.4\msun$ is shown in the top panel as dashes. 
The relative errors of the fits are also shown in the narrow additional panels below the main panels.
}
\label{fig:Mprime}
\end{figure}

\subsection{Connection between basic  parameters of rotating and non-rotating NSs}

The first step in understanding the impact of rapid rotation is to compute the shape of a  NS rotating with a given observed frequency $\nu_*$. 
The most important input parameter is the equatorial radius $R_{\rm e}$ (see Appendix \ref{sec:app1}). 
We would like to find a connection of the radius $R$ and the mass $M$ of a non-rotating NS with the equatorial radius $R_{\rm e}$ and gravitational mass $M'$ of a NS with the same baryonic mass $\bar{M}$, rotating at a rate $\nu_*$. 
\citet{CShT94} computed models of rapidly rotating NSs in general relativity for several EoS and a few values of NS masses, including $1.4\msun$ and the maximum possible NS mass for a given EoS.
The ratio of $R_{\rm e}$  to the radius of the non-rotating configuration $R$ computed from these models is shown in Fig.\,\ref{fig:radius} as a function of  the relative rotation frequency $\bar \nu = {\nu_*}/{\nu_{\rm cr}}$, where   
\be
\nu_{\rm cr} = 1278\,M_{1.4}^{1/2}\, \left(\frac{10~{\rm km}}{R}\right)^{3/2}\ \ \mbox{Hz} 
\ee
is the maximum possible rotation frequency for a given non-rotating NS mass and radius \citep{Haensel.etal:09} and we defined $M_{1.4} = M/1.4\msun$.
The tabulated values can be well represented by an approximate formula 
\be \label{eq:RRe}
 R_{\rm e} =R\,\left[0.9766+\frac{0.025}{1.07-\bar \nu}+\,0.07\,M_{1.4}^{3/2}\,\bar \nu^2\right] ,
\ee
which is shown in Fig.\,\ref{fig:radius} by red curves. 
The relative fitting errors are also shown in the additional panels. 
The maximum errors of fitting using Eq.\,(\ref{eq:RRe}) are not larger than 5\% for the most massive and rapidly rotating NSs. 
We note that we use the extreme EoSs A and L from \citet{CShT94}, which bracket the most probable modern EoS \citep[see e.g.][]{Nattila.etal:16}, giving us a wider range of values to test our formulae.

The mass of the rotating configuration is also affected. 
It is obvious that the gravitational mass of the NS is not a simple sum of the baryon masses $\bar M$, but is reduced roughly as $M=\bar M -G\bar M^2/c^2R$ due to a deep gravitational well.  
Because the radius of the rotating NS is larger than that of a non-rotating NS, the gravitational correction is reduced and the gravitational mass increases.
The ratio of the gravitational mass of a rotating star $M'$ to the mass of a non-rotating star $M$ increases with the rotation rate as shown in Fig.\,\ref{fig:Mprime} based on calculations of \citet{CShT94}. 
The tabulated values can be fitted by a relation (see red solid curves in Fig.\,\ref{fig:Mprime})
\be \label{eq:massprime}
M'=M\left[a_0  + \frac{a_1}{1.1-\bar \nu}+ a_2\,\bar \nu^2\right] ,
\ee
where the fitting coefficients are $a_0=1-a_1/1.1$, $a_1=0.001 M_{1.4}^{3/2}$, and $a_2=10 a_1$. 
The accuracy of Eq.\,(\ref{eq:massprime}) is typically better than 1\% (see additional panels in Fig.\,\ref{fig:Mprime}). 
We also note that relations\,(\ref{eq:RRe}) and (\ref{eq:massprime}) are valid for the NSs of fixed baryonic  mass $\bar{M}$. 
  
\subsection{Apparent area of a rotating NS}
\label{sec:app_area}
 
\begin{figure}
\centering\includegraphics[width=0.75\columnwidth]{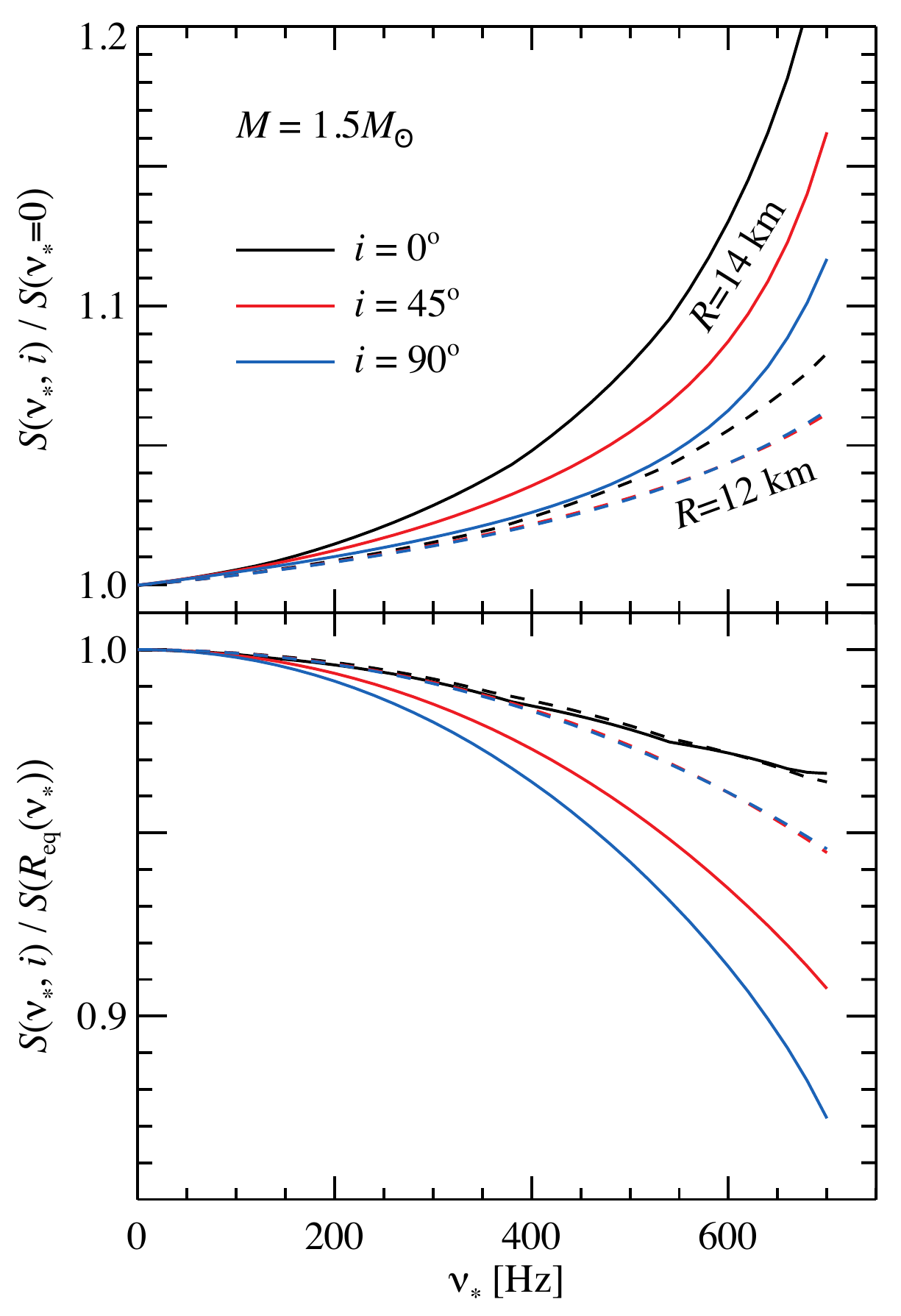}
\caption{{\it Top panel}: Ratio of the apparent area of a rapidly rotating NS to that of the non-rotating star (of the same baryonic mass) as a function of the rotational frequency for inclination angles $i=$0\degr, 45\degr, and 90\degr\ is shown by black, red, and blue curves, respectively.  
The mass of a non-rotating star was assumed $M=1.5\msun$ and two radii $R=14$ (solid curves) and 12 km (dashed curves) are considered. 
{\it Bottom panel}: Apparent area of a rapidly rotating NS for the same family of models as in the top panels divided by the apparent area of a non-rotating NS of the same gravitational mass $M'$ (from Eq.\,(\ref{eq:massprime})) and the radius equal to equatorial radius $R_{\rm e}$ given by Eq.\,(\ref{eq:RRe}). 
The red and the blue dashed curves overlap in both panels.
}
\label{fig:area} 
\end{figure}

The centrifugal force deforms the shape of a rapidly rotating NS, which becomes flattened at the poles and extended at the equator.
This means that the projection area on the sky of a rapidly rotating NS $S(\nu_*,i)$ is different from the visible area of the non-rotating NS of the same baryonic  mass $S(\nu_*=0)=\uppi R^2(1+z)^2$, and it depends on the inclination angle $i$. 
The apparent area of the given family of models of the rapidly rotating NSs of the same baryonic mass is computed via Eq.\,(\ref{eq:FE_oblate}) using the specific intensity
and the energy ratio equal to one, and the approach described further in Appendix\,\ref{sec:app2}. 
The relation between the equatorial radius $R_{\rm e}$ and the radius of the corresponding non-rotating NS is given by Eq.\,(\ref{eq:RRe}), while the masses are related via Eq.\,(\ref{eq:massprime}).
The NS shape is described by approximate formulae from \citet{ALGM:14}, see Appendix\,\ref{sec:app1}.
The relativistic computations of the ratio $S(\nu_*,i) /S(\nu_*=0)$ for a few inclination angles $i$ are presented in the top panel of Fig.\,\ref{fig:area}. 
We note that the apparent area is larger for low inclination and the deformation is smaller for more compact stars. 
Because this ratio is greater than unity, in order to obtain the apparent area of a non-rotating NS, the observed apparent area has to be reduced by a factor depending on the assumed or measured $\nu_*$ and $i$. 

The bottom panel of Fig.\,\ref{fig:area} shows the ratio of the apparent area of a rotating NS to that of a spherical star of the same equatorial radius and same gravitational mass (and thus different baryonic masses). 
This ratio is obviously lower than unity. 
{\citet[][see their Figs.\,5 and 6]{Baubock.etal:12} also presented similar dependences. 
They used a rather different approach to describe rapidly rotating NSs, but their results for the case Kerr + Obl are similar to ours. 
For a better comparison we computed the ratio $S(\nu_*,i)/S(R_{\rm e}(\nu_*))$ for the model with parameters $M' = 1.8 M_\odot$, $R_{\rm e} = 15$\,km, $\nu_*=500$\,Hz (corresponding to $M\approx1.79 M_\odot$ and $R\approx14.13$\,km) and $i = 90\degr$ and got the value of 0.963. 
We note that \citet{Baubock.etal:12} used another approximation for the NS shape from \citet{mors07}. 
Using the same approximation we obtained the value of 0.971, which is very close to the corresponding value of $\approx$0.975 in their Fig.\,5.

\subsection{Emission spectrum of a NS}
 \label{sec:surface_spectra}
 
In the calculations presented in this paper we consider a number of different cases for the local specific intensity. 
The simplest model for the surface emission is that of the blackbody emission, with the specific intensity  measured in the  frame associated with the NS surface being 
\be \label{eq:emit_blackbody}
I'_{E'}(\sigma') = B_{E'}(T) , 
\ee
where $B_{E'}(T)$ is the Planck function of local temperature $T$ at photon energy $E'$ and $\sigma'$ is the zenith angle measured from the local normal to the NS surface. \footnote{Here and below the prime denotes the photon energies and the spectral functions measured at the NS surface, whereas the photon energies and the spectral functions without prime correspond to the values in the observer's frame.}
The corresponding flux in this case is 
\be \label{eq:flux_blackbody}
F'_{E'}  =  2\pi \int_{0}^{\uppi/2} B_{E'}(T) \cos\sigma'\, \sin\sigma'\, d\sigma' = \pi B_{E'}(T)  
\ee
and the bolometric flux corresponding to this case is given by the Stefan--Boltzmann law $F_{\rm bol}=\sigma_{\rm SB}T^4$.  
Another case for the angular distribution that we would like to explore is that corresponding to the electron-scattering dominated atmosphere. 
We keep the Planck function to describe the energy dependence of the local specific intensity and approximate the dependence on the zenith angle by a linear function of $\cos\sigma'$ \citep{ChaBre47, sob49}: 
\be \label{eq:emit_elsc}
I'_{E'}(\sigma') = B_{E'}(T)\  (0.421+0.868\,\cos \sigma').
\ee 
The chosen normalization of the angular factor gives the same emergent flux as a unit constant.

Furthermore, we consider the case of a diluted blackbody corresponding to the intensity 
\be \label{eq:emit_diluted_blackbody}
I'_{E'}(\sigma') = w B_{E'}(\fc \teff) , 
\ee 
where $w$ is the dilution factor, $\fc$ is the colour-correction factor,  $\teff$ is the effective temperature, and $T_{\rm c}=\fc \teff$ is the colour temperature. 
The surface flux is now 
\be \label{eq:flux_diluted_blackbody}
F'_{E'}  = w\, \pi B_{E'}(\fc \teff)  .
\ee
Similarly to Eq. (\ref{eq:emit_elsc}) we consider a diluted blackbody with a beamed emission pattern 
\be \label{eq:emit_diluted_elsc}
I'_{E'}(\sigma') = w\, B_{E'}(\fc \teff) \  (0.421+0.868\,\cos \sigma').
\ee 
The surface flux is also given by Eq. (\ref{eq:flux_diluted_blackbody}). 
Equation (\ref{eq:emit_diluted_elsc}) provides a good approximation to the specific intensity from the NS atmosphere models computed by \citet{SPW12, Suleimanov.etal:17} for NS luminosities close enough to the Eddington value ($L \gtrsim  0.1 \ledd$).  
These accurate model for the intensity for various effective temperatures $\teff$ and gravities $g_{\rm eff}$ are also used in our calculations. 
They are also fitted by Eq. (\ref{eq:emit_diluted_elsc}) with free $w$ and $\fc$, which are tabulated on a grid of $\teff$ and  $g_{\rm eff}$ and used in the calculations instead of accurate  models for simplicity.

The flux from a bursting NS as observed at Earth can also be well represented by a diluted blackbody: 
\be \label{eq:obs_blackbody}
F_E = K\ \pi B_E(T_{\rm BB}). 
\ee
Here the normalization factor $K$ is proportional to the dilution factor $w$, while the observed colour temperature $T_{\rm BB}$ is related to the colour-temperature $\fc\teff$ as measured at the NS surface.

\begin{figure*}
\centering
\includegraphics[width=.75\columnwidth]{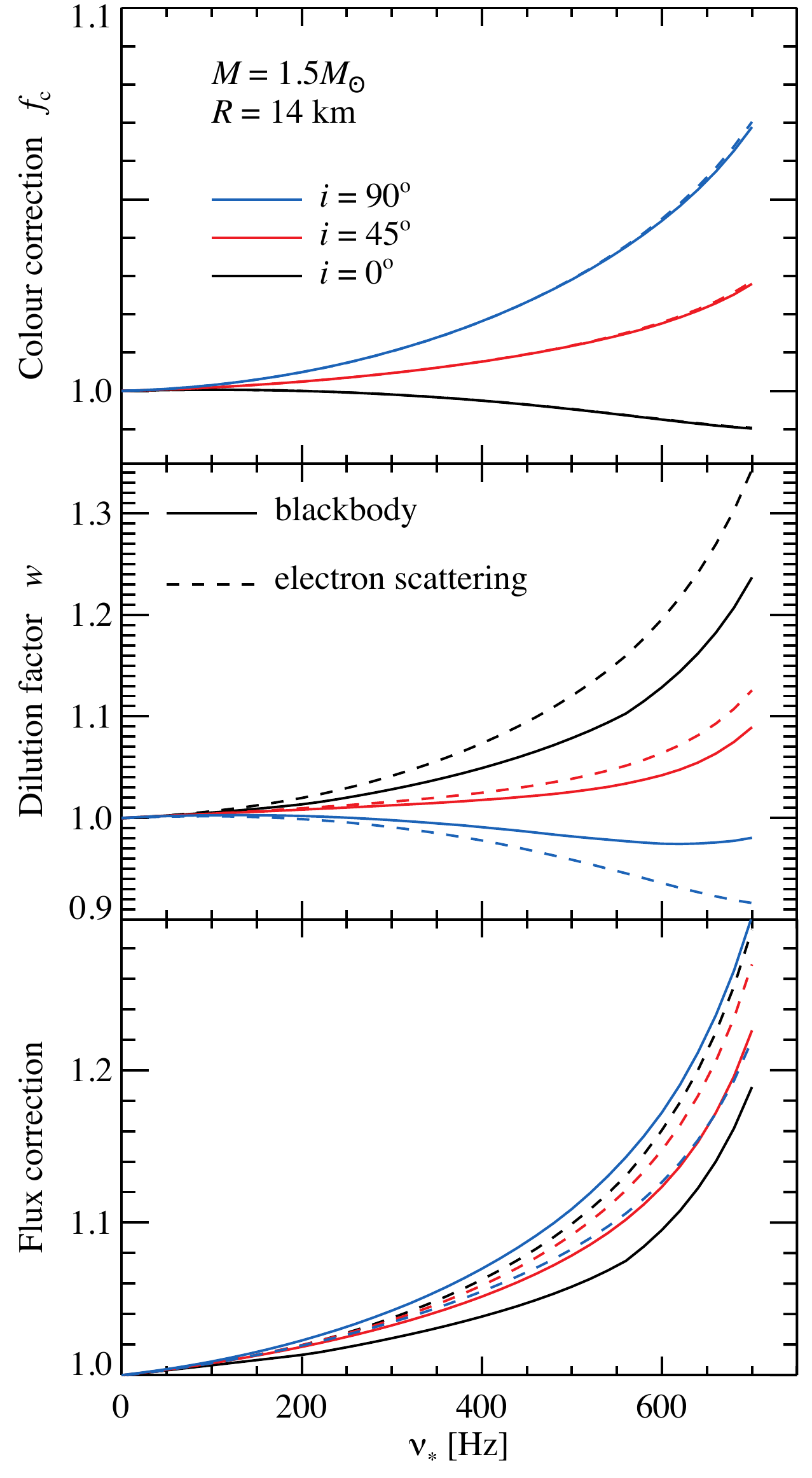}
\hspace{1cm}
\includegraphics[width=.75\columnwidth]{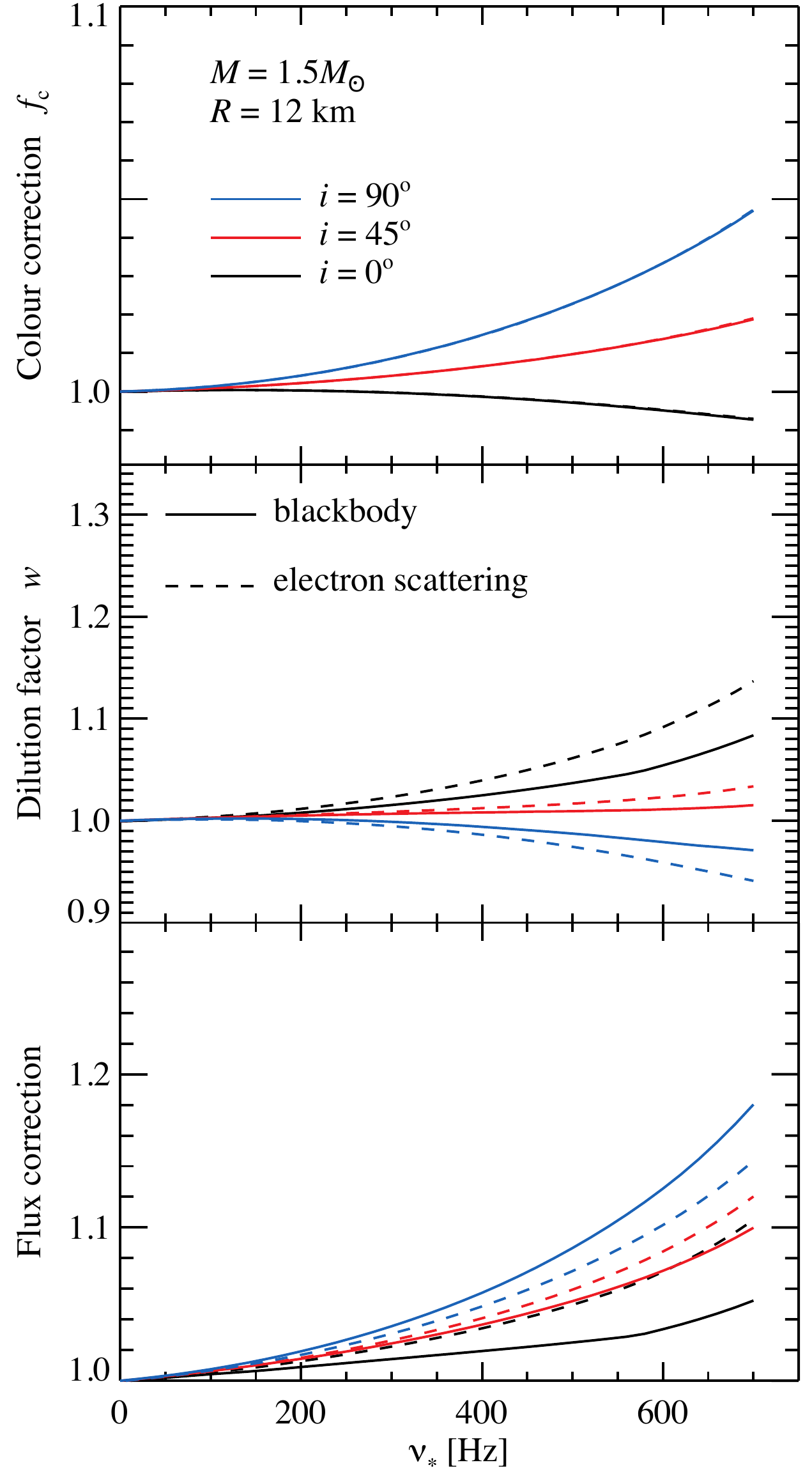}
\caption{{\it Top panels}: Colour correction to the observed blackbody temperature compared to a non-rotating NS for a NS of constant local temperature 
as a function of the rotational frequency for  inclination angles $i=$0\degr, 45\degr, and 90\degr\ is shown by the black, red, and blue curves, respectively.   
The left panels correspond to radius of a non-rotating NS $R$=14~km and the right panels to $R$=12~km. 
The mass of the non-rotating NS was assumed $M=1.5\msun$ and both families of models have the same baryonic mass. 
The solid curves are for emission pattern corresponding to the isotropic intensity (\ref{eq:emit_blackbody}), while the dashed curves are for the electron-scattering dominated atmosphere with the intensity given by Eq. (\ref{eq:emit_elsc}).  
The colour-correction factor is indistinguishable for the two considered emission patterns. 
{\it Middle panels}:  Correction factor to the blackbody normalization (dilution factor) as a function of the rotational frequency. 
{\it Bottom panels}: The correction to the bolometric flux. 
Same notations as in the top panels. 
}
\label{fig:corrections} 
\end{figure*}

\subsection{Apparent spectrum of a blackbody emitting rotating NS}
 \label{sec:app_spectra}

The apparent size of the NS on the sky does not fully determine how the spectral parameters change due to rotation. 
Let us assume that the NS surface emits locally a blackbody of constant local temperature $T= T_{\rm c} = \teff$ with specific intensity given by Eq. (\ref{eq:emit_blackbody}). 
The observed spectrum for a rotating NS deviates from the blackbody for two reasons. 
Firstly, the surface gravitational redshift varies over the surface of a rotating NS. 
Secondly, Doppler boosting affects the amplitude and energy of the spectral peak. 
The Doppler boost is insignificant for pole-on sources and reaches its maximum level  at $i=90\degr$.  
The change in the spectrum thus depends not only on the NS rotation rate which affects its shape, but also on the inclination angle $i$. 

Let us construct families of rotating NS models which have the same baryonic mass, but have different rotational frequencies. 
We first choose the mass $M=1.5\msun$ and consider two different radii $R=14$\,km and 12~km for a non-rotating star.
Then for each spin, the values for the rotating NS mass $M'$ and equatorial radius $R_{\rm e}$ were computed using Eqs.\,(\ref{eq:RRe}) and (\ref{eq:massprime}).
The observed spectrum is computed following the approach described in Appendix \ref{sec:app2}. 
The resulting spectrum $F_E$ in the frame of the observer at infinity was fitted in the observed energy range (0.1--10)$kT$ with the diluted blackbody 
$w\uppi B_E(\fc T_{\infty})$ giving the values of $\fc$ and $w$. 
Here  $T_{\infty} = T/(1+z)$ is the gravitationally redshifted local temperature, where $z$ is the gravitational redshift of the non-rotating NS, for which $\fc=1$. 
The dilution factor $w$ was then divided by the value obtained for a non-rotating NS.

\begin{figure*}
\centering
\includegraphics[width=0.58\columnwidth]{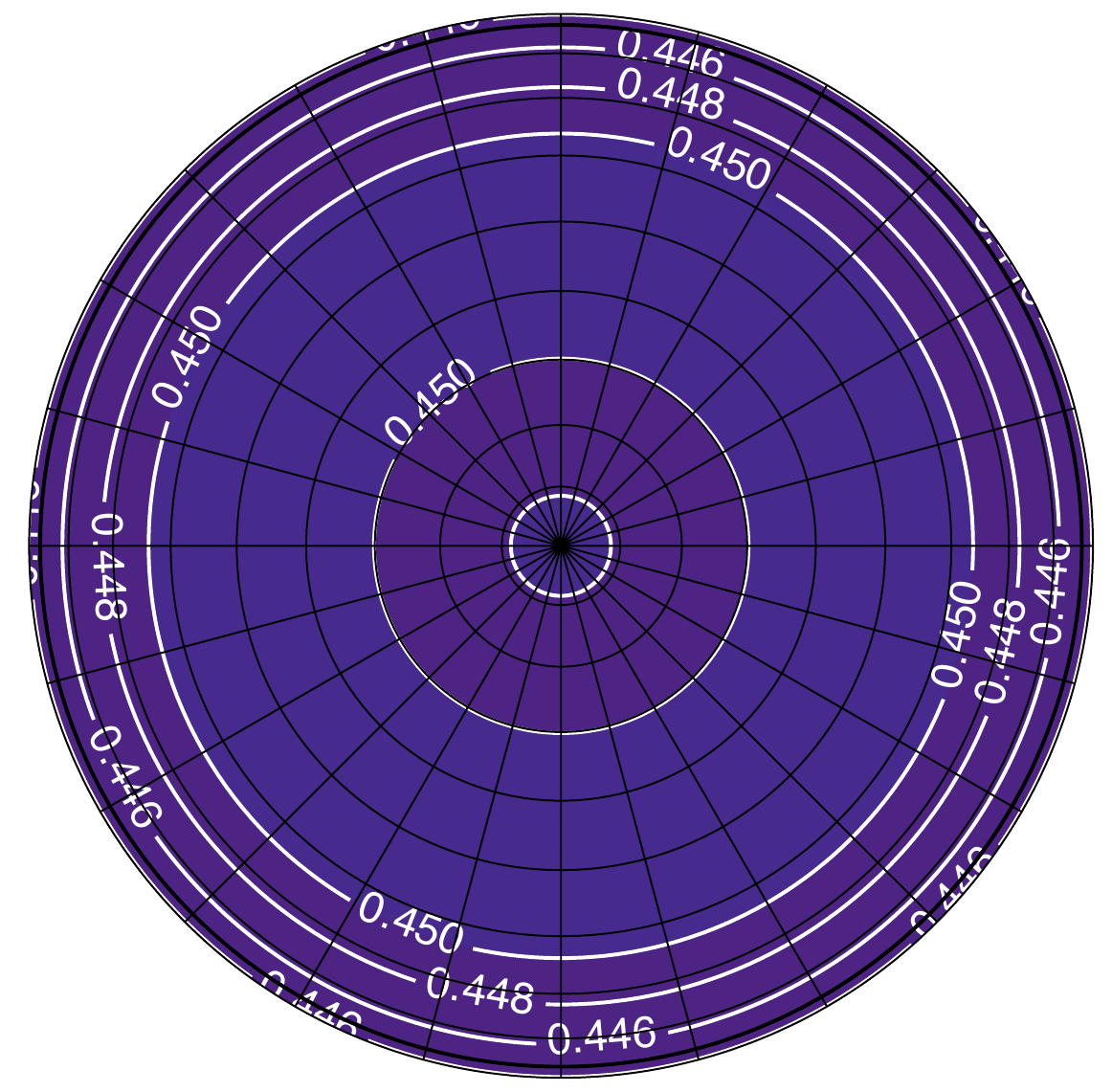} 
\includegraphics[width=0.6\columnwidth]{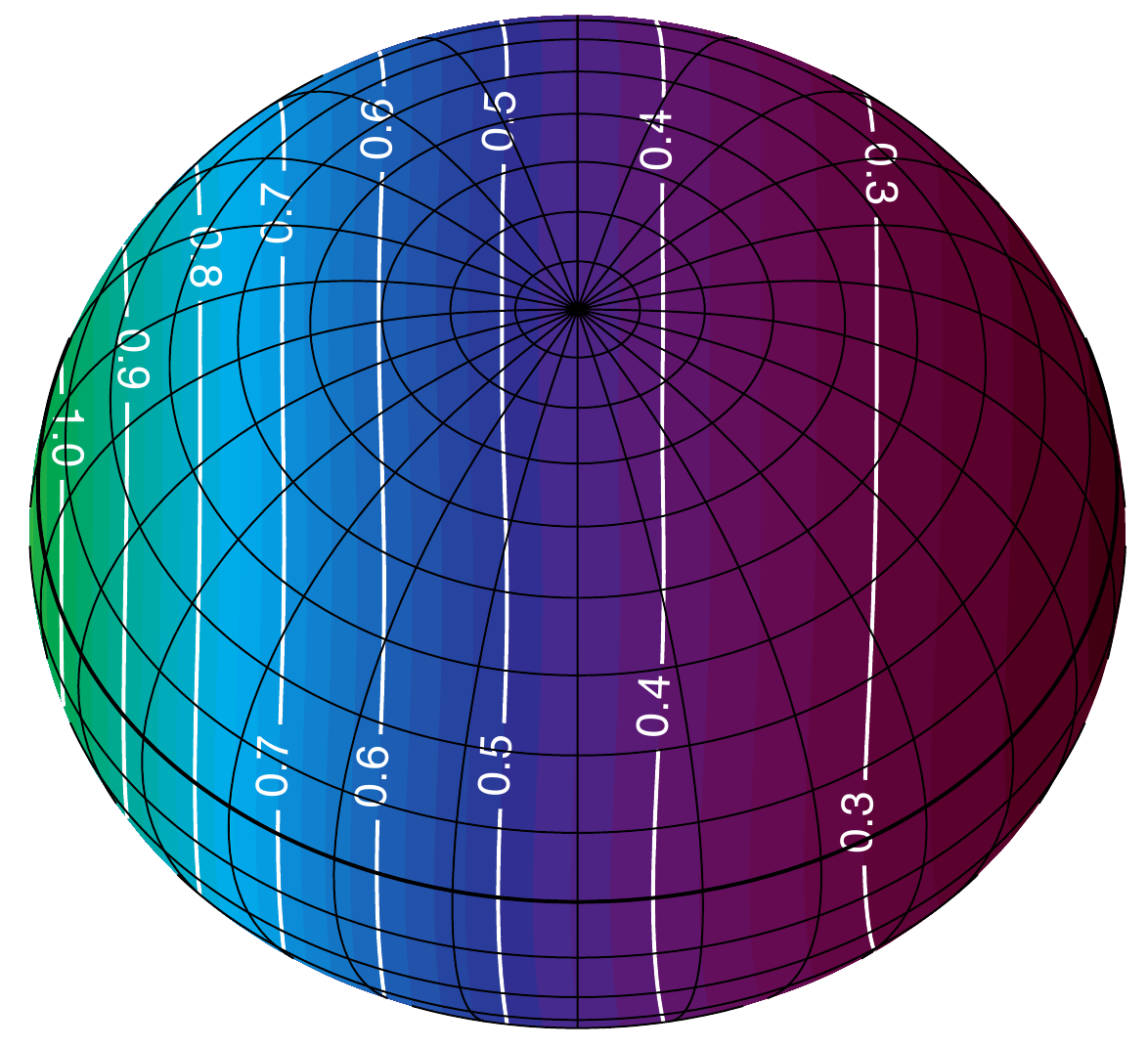} 
\includegraphics[width=0.6\columnwidth]{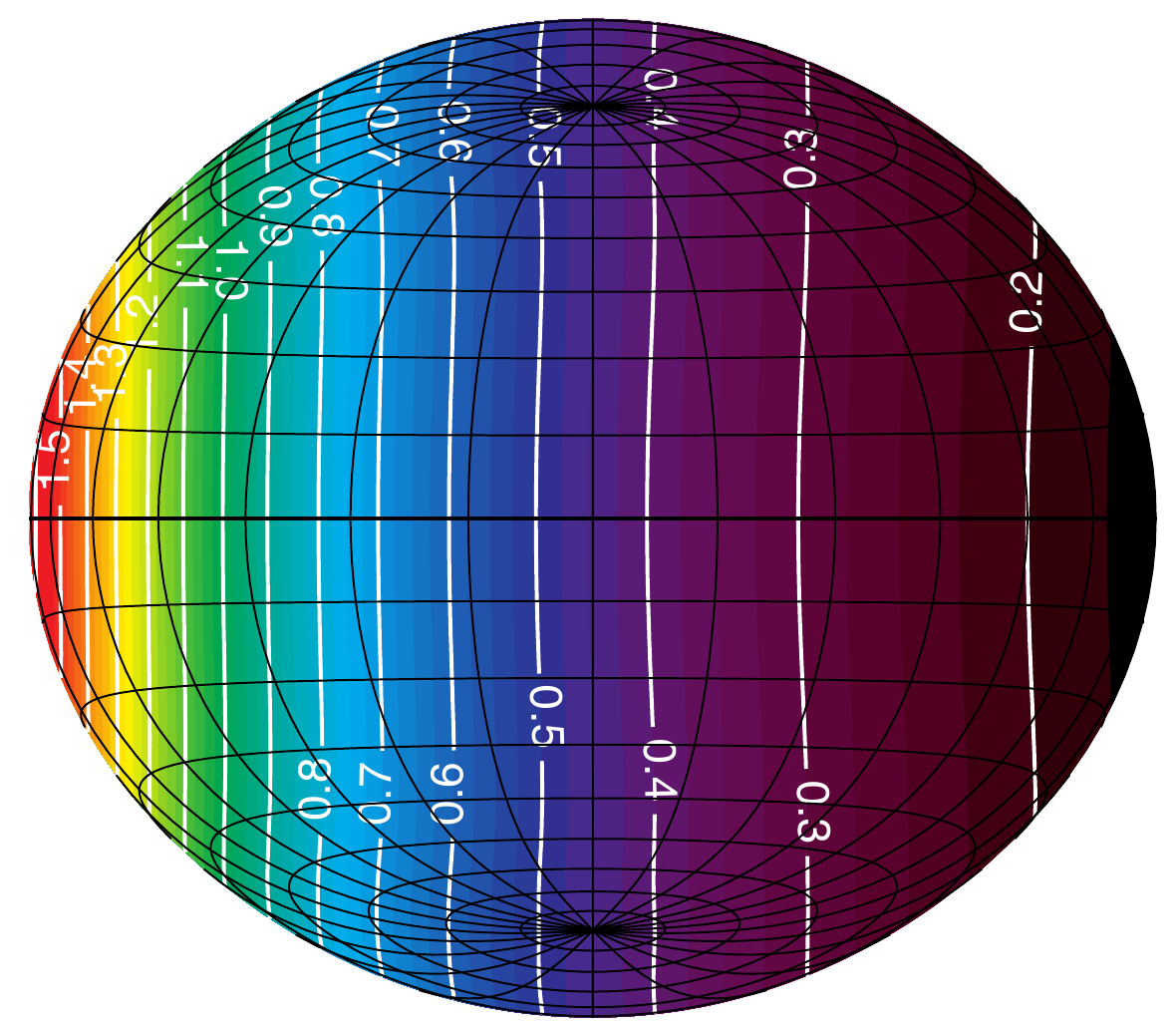} 

\includegraphics[width=0.6\columnwidth]{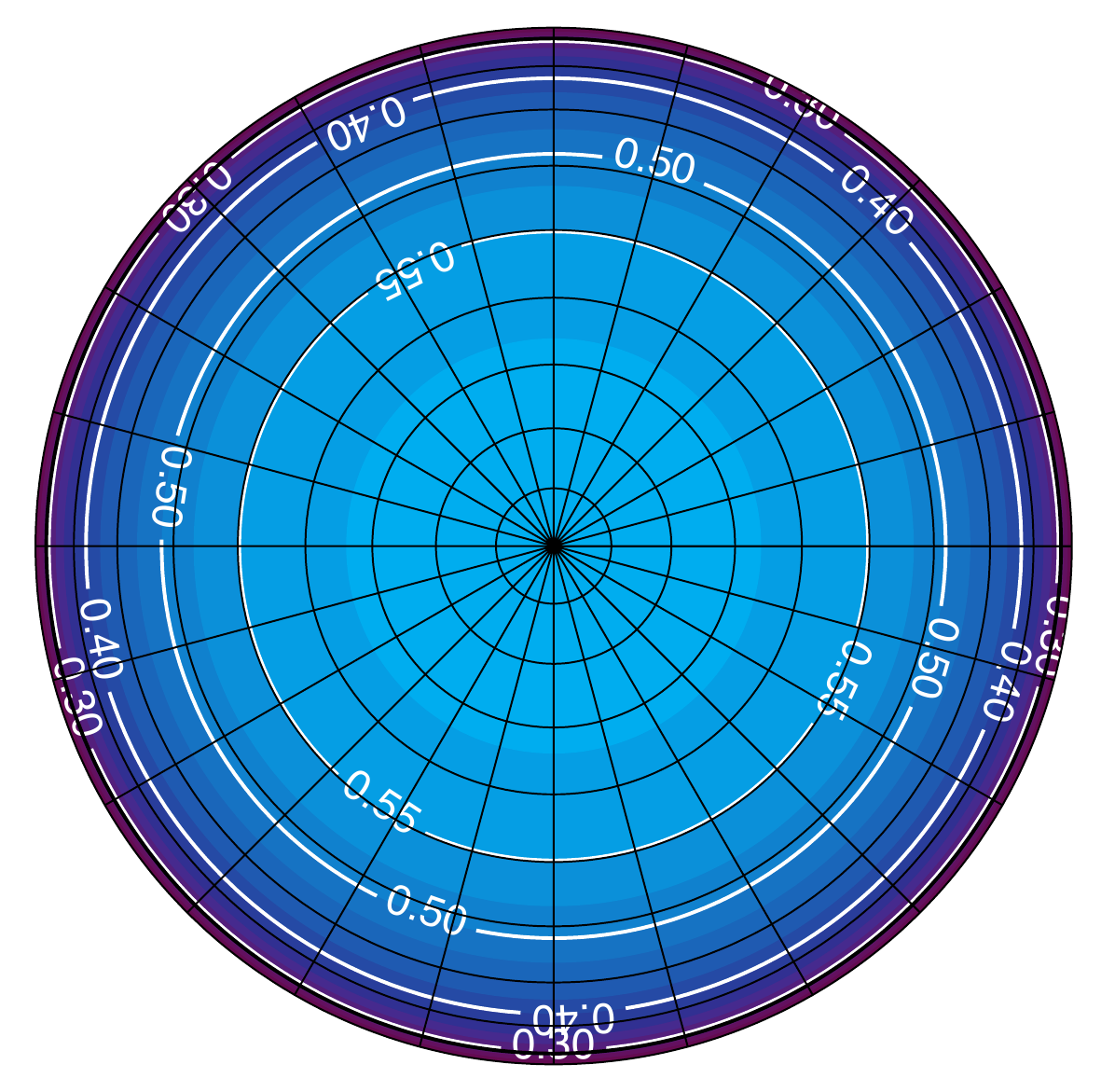} 
\includegraphics[width=0.6\columnwidth]{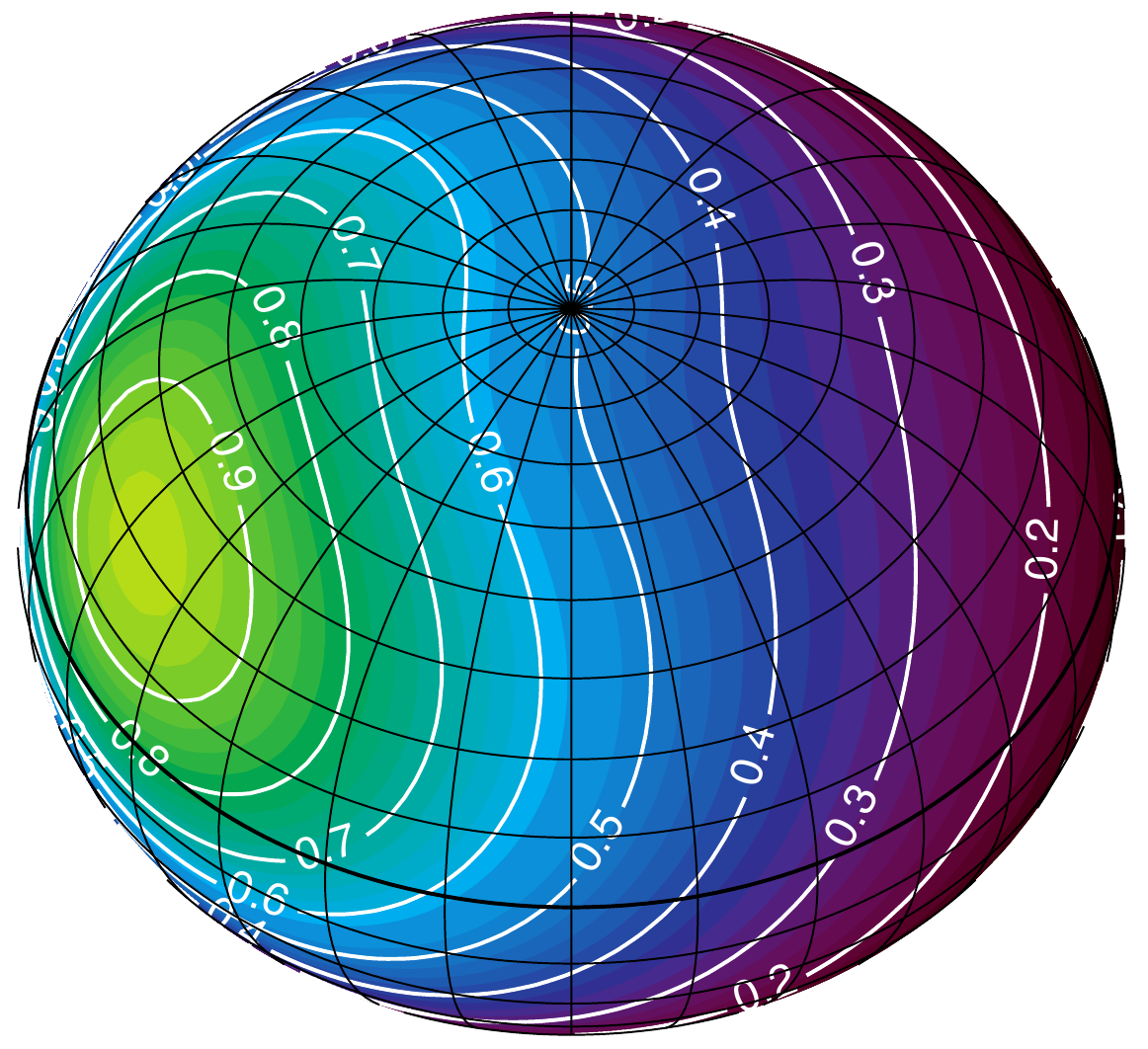}
\includegraphics[width=0.6\columnwidth]{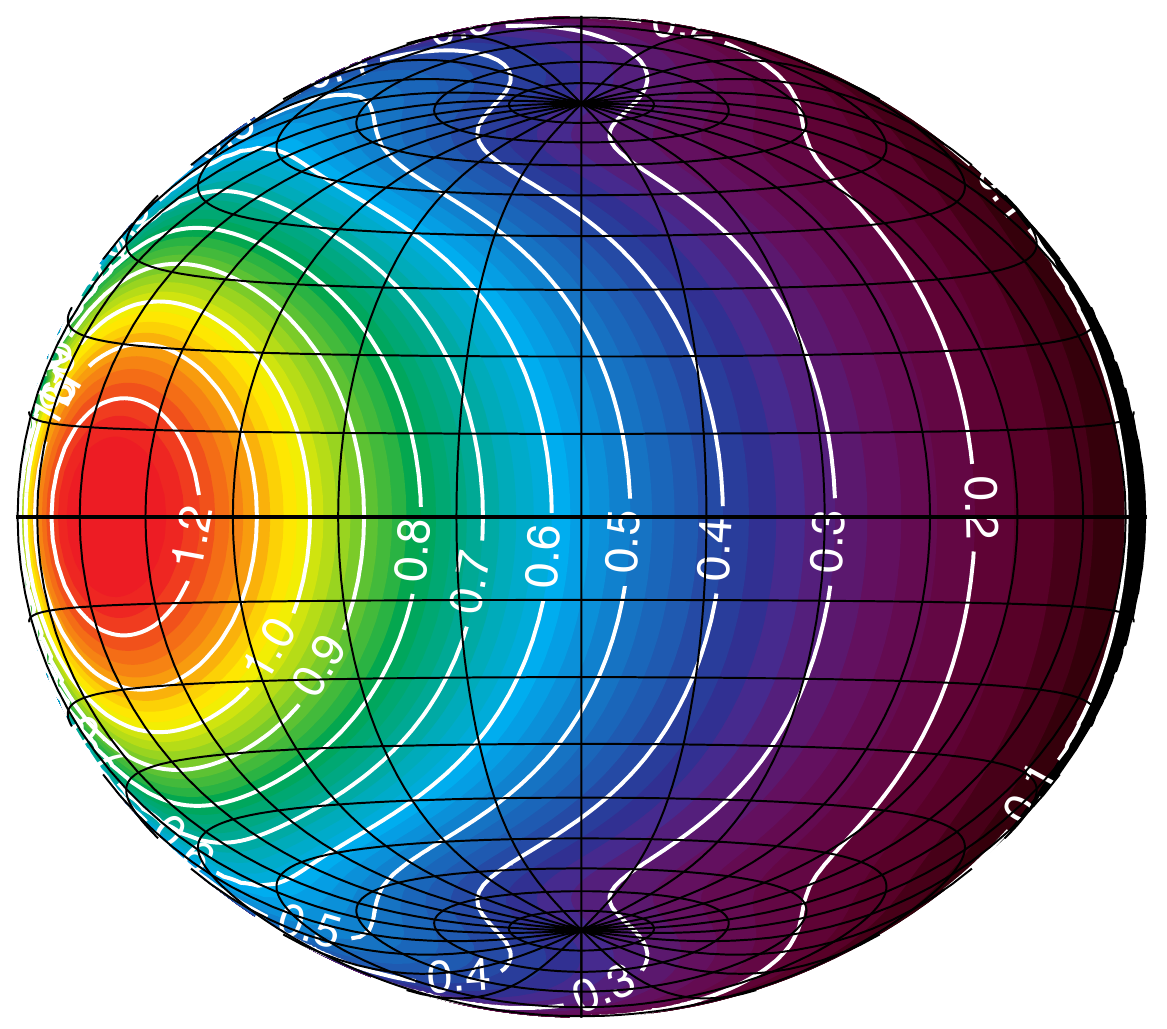} 
\caption{Images of a NS rotating at 700 Hz. 
The mass and  the radius of a non-rotating NS of the same baryonic mass are $M=1.5\msun$ and $R=14$ km.  
The white curves and the colours show the contours of constant bolometric intensity. 
\textit{Upper panels}: Blackbody case of local isotropic intensity. 
\textit{Lower panels}: Electron-scattering dominated atmosphere. 
Panels from the left to the right correspond to the inclinations $i=0\degr$, 45\degr, and 90\degr. 
The lines of constant latitudes (every 10\degr) and longitudes (every 15\degr) are shown in black.  
The colour scheme is different for the two rows. }
\label{fig:images_NS}
\end{figure*}

We plot the corresponding colour-correction factor $\fc$, correction to the dilution factor $w$, and the correction to the bolometric flux with the solid curves in Fig.\,\ref{fig:corrections}. 
The top panels demonstrate that the colour temperature decreases  with the spin for a face-on observer ($i=0$) because  the NS oblateness increases at higher rotation rates resulting in a corresponding increase in the gravitational redshift of the polar region. 
On the other hand, the edge-on observer sees higher temperature comparing to a non-rotating star because of the Doppler effect. 
The effect is greater for larger NS radii. 

The middle panels of Fig.\,\ref{fig:corrections} show how the blackbody normalization is affected by rotation. 
We see that for low inclinations the normalization increases, which results from an increase in the apparent area as discussed in Sect.\,\ref{sec:app_area}. 
However, for high inclinations  the normalization decreases slightly, which is opposite to the behaviour of the apparent area. 
This is a natural consequence of a large increase in $\fc$ because of the Doppler effect. 
However, the bolometric flux (shown in bottom panels of Fig.\,\ref{fig:corrections}) increases a little more slowly than $\fc^4$ resulting in a reduction of $w$.
The bolometric flux monotonically increases with the inclination being highest for an edge-on observer. 
It follows very closely the product $w\fc^4$, with the deviation from that dependence (i.e. bolometric correction to the blackbody) being negligible for $i=0\degr$ and reaching only 1--2\% for $i=90\degr$ at $\nu_*=700$\,Hz.

In addition to the blackbody, we also considered the case of an electron-scattering dominated atmosphere, where the intensity given by Eq. (\ref{eq:emit_elsc}) is much stronger beamed along the surface normal.
The results are shown with dashed curves in Fig.\,\ref{fig:corrections}.
Interestingly,  the colour-correction factor turned out to be identical to that obtained for isotropic emission. 
The correction to the blackbody normalization, on the other hand, deviates farther from unity  than for the isotropic emission (see middle panels of Fig.\,\ref{fig:corrections}).  
Also in this case, the bolometric flux increases with the NS spin.
For smaller NS radii, $R=12$\,km (right bottom panel), its dependence on the inclination angle is similar to the isotropic case, i.e. it monotonically increases with inclination. 
However, for larger radii, $R=14$\,km, it shows the opposite behaviour, with the flux being highest for a face-on observer. We note that the dashed curve below the blue solid curve in the left bottom panel of Fig.\,\ref{fig:corrections} is black and corresponds to $i=0\degr$.
The reason is a stronger radiation beaming along the normal and greater deformation of the NS surface.
The increase in the apparent area for a face-on observer makes a stronger effect than the increase in  the Doppler boost for an edge-on observer.  

At this point we can try to compare our results with those of \citet{Baubock.etal:15}. 
However, there is a large difference in our approaches that complicates direct comparison.
\citet{Baubock.etal:15} computed the corrections to the colour temperature and bolometric flux too, but compared the results obtained for a rotating star with equatorial radius $R_{\rm e}$ to those for a non-rotating star of the same radius.
In this case the overall gravitational redshift is similar for the two stars, and the colour temperature changes mostly due to the Doppler effect resulting in a 1\% deviation. 
However, we  fix the mass and radius of a non-rotating star and consider a family of models as a function of spin. 
Thus, at high spin rates, our $R_{\rm e}$ is significantly larger and polar radius is significantly smaller than the radius $R$ of a non-rotating configuration. 
For low inclinations, this results in an effectively higher gravitational redshift leading to a decrease in the colour temperature (and $\fc$) by 1\% compared to a non-rotating case. 
On the other hand, at high inclination, the effective gravitational redshift is much lower resulting 
in an observed temperature that is higher by 7\% for a rotating star.
We also note  that corrections to the metric due to the frame dragging and quadrupole moment play no role for spins below  500 Hz.
 
As an illustration we also show in Fig.\,\ref{fig:images_NS} the images of NSs rotating  at 700 Hz (see also \citealt{Vincent.etal:18}). 
The colour-coding and the white lines are the contours of constant observed  bolometric  intensity.  
It is equal to the product of the intensity as measured  at the NS surface $I(\sigma')$ and the fourth power of the total redshift (i.e. the product of Doppler factor and the surface gravitational redshift, given by Eq. \ref{eq:EEpr_drag}). 
We  consider the same two cases of an isotropically emitting star (see upper panels of  Fig.\,\ref{fig:images_NS}) and  an electron-scattering dominated atmosphere (lower panels of  Fig.\,\ref{fig:images_NS}). 
The results are normalized by the factor $\sigma_{\rm SB}T^4/\pi$. 
The face-on image of an isotropically emitting star (left upper panel) shows the minimum at the pole because the redshift there is highest. 
The intensity is also low at the equator because of the transverse Doppler effect.
At high inclinations, the contours of constant intensity nearly follow the lines of constant surface radial (projected) velocity which are nearly straight vertical lines. 
The highest and the lowest intensities are reached at the edges of the image, where the absolute values of the projected velocities are highest. 
The stronger beaming of radiation along the normal to the surface in the second case dominates over the reduction of intensity due to higher redshift, resulting in a peak of emission at the pole for a face-on star (left lower panel). 
At higher inclinations, the lines of constant intensity are no longer straight because of the strong influence of beaming. 
Along the vertical line the angle $\sigma'$ at which we see the surface element is different, resulting in different local intensity.
Thus, the observed intensity peak is now not at the edge (where $\sigma'=\uppi/2$), but is reached at the point where the angle of the line of sight to the local normal is smaller. 
We note here that our computation method is very efficient: computing one image of resolution $70\times400$ in latitude and azimuth on a standard laptop with {\sc idl} takes less than 0.3\,s, which is about 3000 times faster than would be needed if an exact ray-tracing code is used \citep[e.g.][]{Vincent.etal:18}. 

\section{Modified direct cooling tail method}
\label{sec:method}
 
\subsection{Non-rotating X-ray bursting neutron stars}
\label{sec:directcooling}

The direct cooling method \citep{Suleimanov.etal:17} assumes that the investigated NS with given mass $M$ and radius $R$  rotates slowly, its shape is spherical, and the surface has a uniform effective temperature $\teff$. 
 In this case a single model atmosphere with the given $\teff$, the surface gravity $g=GM(1+z)/R^2$ and the chemical composition is enough to describe the spectrum of the whole NS. 
The gravitational redshift factor is related to the NS Schwarzschild radius $R_{\rm S} = 2GM/c^{2}$ as $1+z = (1-R_{\rm S}/R)^{-1/2}$.   
Therefore, we can connect the local atmosphere parameter $\teff$ with the whole observed NS luminosity  $L=4\uppi R^2\,\sigma_{\rm SB} \teff^4/(1+z)^{2}$, and use the relative NS luminosity $\ell=L/\ledd$ instead of $\teff$. 
Here 
\be 
\ledd= \frac{4\uppi GMc}{\kappa_{\rm e}(1+z)}
\ee 
is the NS Eddington luminosity,  $\kappa_{\rm e}=0.2(1+X)$\,cm$^2$\,g$^{-1}$  is the Thomson opacity, and $X$ is the hydrogen mass fraction in the atmosphere. 

It is well known \citep{London86, Lapidus.etal:86} that the local spectra of hot NS model atmospheres $F'_{E'}$ are  well described with a diluted blackbody  (\ref{eq:flux_diluted_blackbody}), where the dilution and the colour-correction factors approximately follow the relation $w\approx \fc^{-4}$ and $\fc > 1$.  
This  explains why the observed spectra of X-ray bursting NSs are usually very well fitted with a blackbody given by Eq. (\ref{eq:obs_blackbody}) 
\citep[see e.g.][]{Galloway08}, 
where $T_{\rm BB} = \fc \teff/(1+z)$ is the observed colour temperature.
As a result, the blackbody normalization parameter $K$ used in the fits depends on the dilution factor $w$ \citep[see details in][]{SPRW11}
\be \label{kbb}
  K = w\, \Omega , 
\ee
where $D$ is the distance to the source and 
\be 
\Omega= \left[\frac{R(1+z)}{D}\right]^2
\ee 
is a geometrical dilution factor proportional to the solid angle occupied by the NS on the sky. 

The parameters $w$ and $\fc$ are found by fitting the emergent model spectra of hot NS atmospheres in the energy band of some X-ray instruments, like \textit{RXTE}/PCA, blueshifted to the NS surface.  
The values of these parameters depend mainly on the relative luminosity $\ell$, although they depend on the surface gravity $g$ and the chemical composition as well. 
We have computed an extended grid of models \citep{SPW11,SPW12,Nattila.etal:15} for a range of $\log g$  from 13.7 to 14.9 with the step 0.15 and for various atmosphere chemical compositions (pure hydrogen, pure helium, solar abundance, solar H/He mix with reduced and increased heavy element abundances, and pure iron).
The model spectra were fitted with a blackbody resulting in dependences $w - \ell$ and $\fc - \ell$.

The direct cooling tail method is based on the results of these computations. 
It is possible to demonstrate \citep{Suleimanov.etal:17} that  the observed dependence 
$K - F_{\rm BB}$ during the thermonuclear burst cooling tail has to be fitted with the model
dependence $w - w\fc^4\,\ell$ computed for each pair of parameters $(M, R)$ and the actual 
chemical composition of the atmosphere.\footnote{The model closely fits the data only for the 
bursts taking place during hard spectral persistent states of LMXBs 
\citep{SPRW11, Poutanen.etal:14, Kajava.etal:14} because of the strong influence 
of accretion during the soft states.}   
Formally, there are two fitting parameters $\Omega$ and the observed Eddington flux 
\be 
F_{\rm Edd,\infty} = \frac{\ledd}{4\uppi D^2} = \frac{GMc}{\kappa_{\rm e}D^{2}\,(1+z)}, 
\ee 
but actually for every pair $(M, R)$ they both depend   on the distance to the source $D$ alone because the $z$ value is known. 
The fitting procedure provides a $\chi^2$ map, which can be used to estimate the most probable values for the NS mass and radius \citep[see examples in][]{Sul.etal:17,Suleimanov.etal:17}.

\begin{figure}
\centering\includegraphics[width=.7\columnwidth]{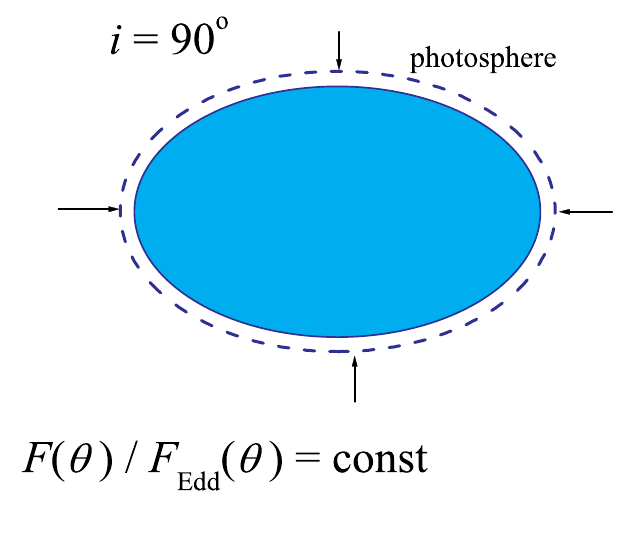}
\includegraphics[width=.78\columnwidth]{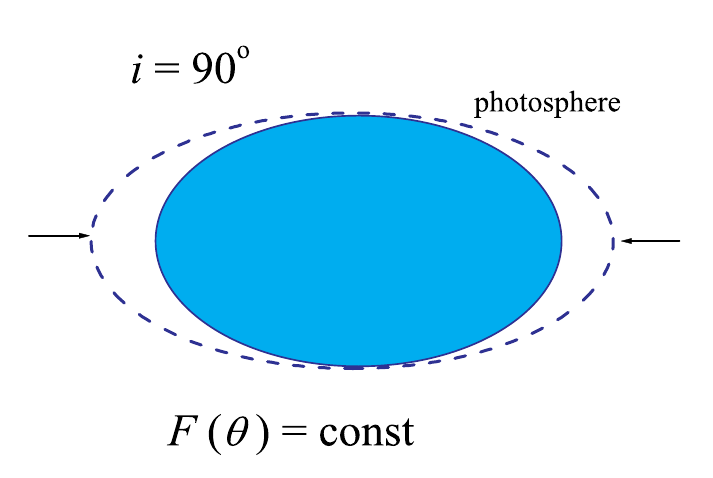} 
\caption{Qualitative difference between the shapes of the extended photospheres (dashed contours) at the touch-down point for two different hypothesis on the latitudinal distribution of the radiation flux  over the surface of a rapidly rotating NS, 
CRF case (top panel) and CAF case (bottom panel).
}
\label{fig:shapephoto}
\end{figure}

\subsection{Spectra of bursting rapidly rotating NSs}
\label{sec:app_burst}

Once we understand the observational appearance of  the blackbody emitting NS as discussed in Sect.\,\ref{sec:rotatingNS}, we  can discuss the emission of a bursting rotating NS. 
One of the effects that needs to be accounted for is the change in the effective gravity along the latitude.
The effective gravity can be presented approximately as $g_{\rm eff} = g - \Omega_*^2 R(\theta) \sin^2 \theta$, where $R(\theta)$ is the NS radius at 
co-latitude  $\theta$. 
The approximate expressions of the effective surface gravity using relativistic computations  were suggested by \citet{ALGM:14} (see also Appendix \ref{sec:app1}).  
It is clear that the effective surface gravity at the poles is stronger than at the equator. 
It means that the value of the local Eddington flux depends on latitude. 
This also implies that it is not possible to unambiguously determine the touch-down point in PRE bursts because it is not known a priori how the actual radiation flux is distributed over the NS surface. 
We can only make some suggestions. 
We consider two cases.  
The first option is that the bolometric flux in units of the local Eddington flux $F_{\rm Edd} = c\,g_{\rm eff}/\kappa_{\rm e}$  is constant over the NS surface (i.e. $F(\theta)/F_{\rm Edd}(\theta)=\ell$). 
In this constant relative flux (CRF) case the  photosphere of the extended envelope touches the NS surface simultaneously and we observe this moment as a touch-down point in the X-ray burst light curve. 
The second case assumes constant absolute flux (CAF), $F(\theta) =$\,const, at every co-latitude equal to the flux of a non-rotating NS with a given relative luminosity $\ell$.
In this case, the photosphere touches the NS surface at the poles first and we can associate the touch-down point with this moment  (see Fig.\,\ref{fig:shapephoto}).  
The real latitudinal flux distribution is not known, and we can only guess that it lies between the two limiting cases considered above.  
In this work we consider both possibilities as equiprobable and estimate the effect of rapid rotation on the observed spectra and derived NS parameters for both cases.  
We note here that constant surface flux does not mean same spectrum of the emitted radiation, because the real atmosphere spectrum depends not only on the effective temperature, but also on local gravity (i.e. the colour-correction factor varies along the latitude). 
Thus, even this problem cannot be reduced directly to the case of a blackbody emitting NS considered in Sect.\,\ref{sec:app_spectra}.

\begin{figure}
\centering
\includegraphics[width=0.9\columnwidth]{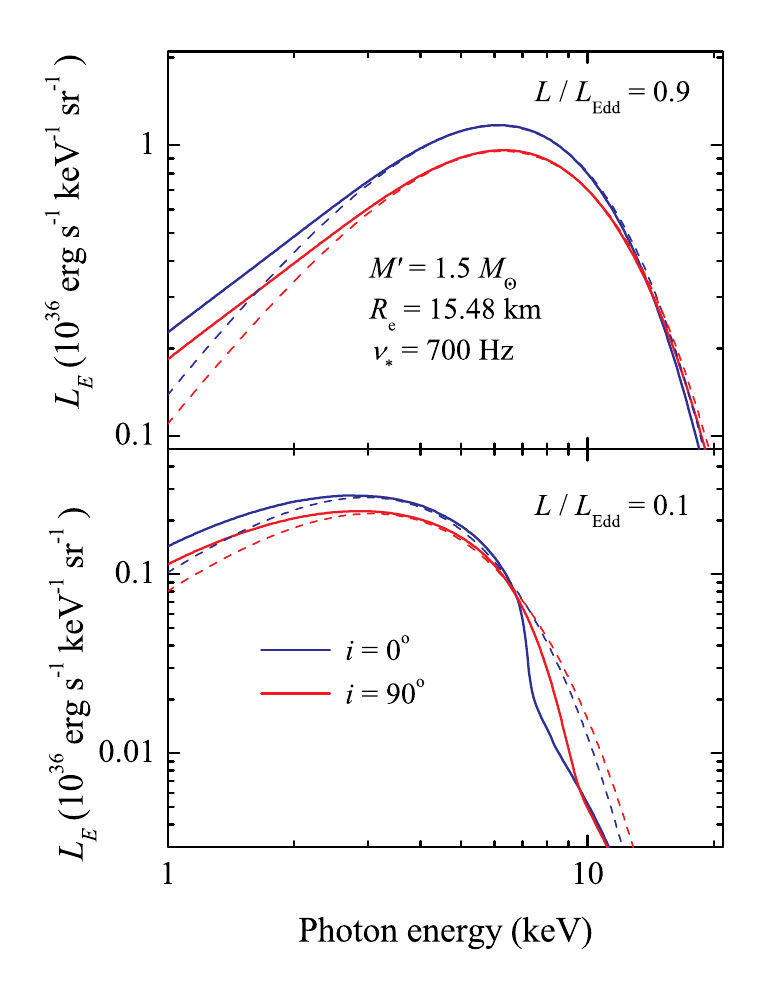}
\caption{Spectra of the rotating NS with the parameters $M'= 1.519\,\msun$, $R_{\rm e}=15.48$\,km, and $\nu_*=700$\,Hz computed for two inclination angles $i=$\,0\degr\, (blue solid curve) and $i=$\,90\degr\ (red solid curves), and for two 
relative luminosities (same over the surface) $\ell=$\,0.9 ({\it top panel}) and $\ell=$\,0.1 ({\it bottom panel}). 
The solid curves represent spectra computed taking accurate local spectra for atmosphere models from \citet{SPW12}. 
The dashed curves correspond to the diluted blackbody approximation of the local spectra. 
Solar chemical composition of the atmosphere is assumed. 
 }
\label{fig:spectra}
\end{figure}

\begin{figure}
\centering{\includegraphics[width=0.9\columnwidth]{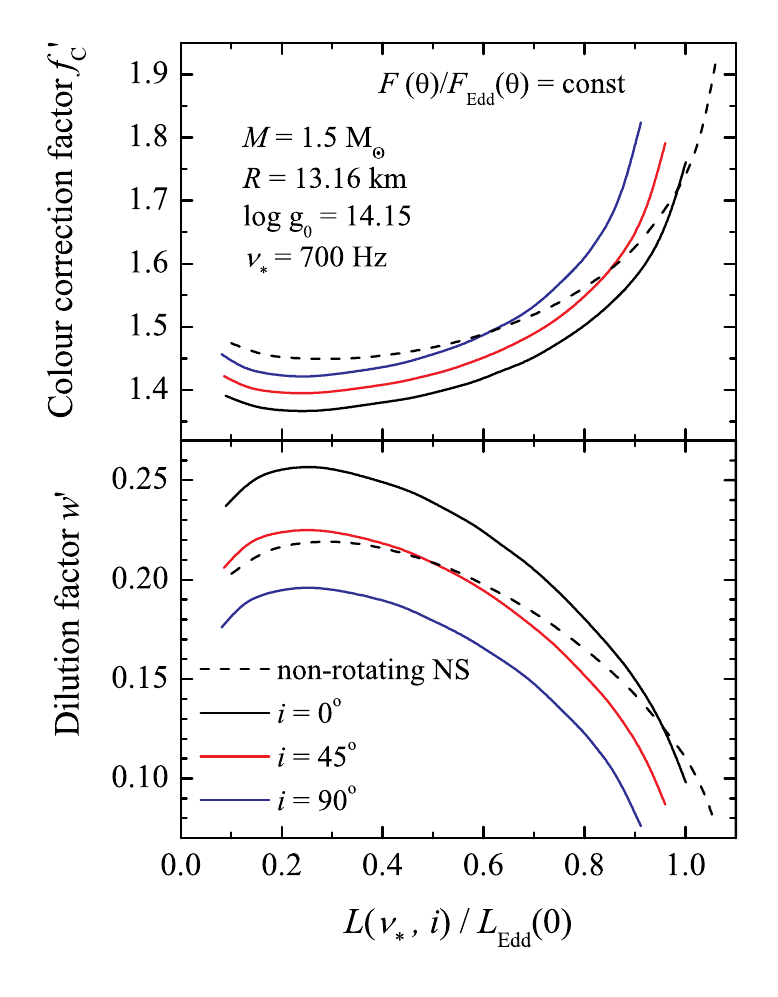}}
 \caption{Dependences of $\fc'$ and $w'$ on the relative luminosity for a rapidly rotating NS for three inclination angles.
The corresponding dependences for a non-rotating NS with the surface gravity $\log g_0 =$\,14.15 are shown with the dashed curves. 
Here chemical composition is solar with the heavy element abundances reduced by 100 times. 
The constant relative flux distribution over the NS surface is assumed (CRF case).
}
\label{fig:fcw_relative}
\end{figure}

\begin{figure}
\centering{\includegraphics[width=0.9\columnwidth]{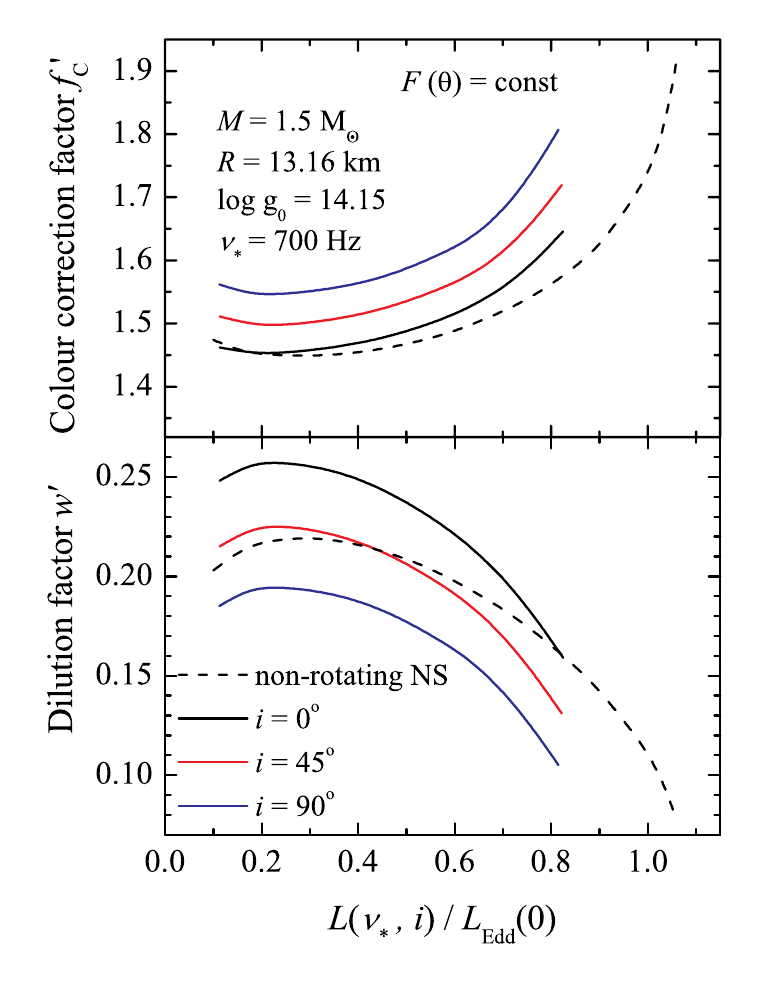}}
 \caption{Same as Fig.\,\ref{fig:fcw_relative}, but computed with constant absolute flux distribution over the NS surface (CAF case).
 }
\label{fig:fcw_absolute}
\end{figure}

Now let us discuss the  appearance of the bursting rotating NS to a distant observer at inclination $i$ for a given relative luminosity $\ell$.
The details of the method to compute the observed spectrum  are presented in Appendix \ref{sec:app2}.
The local specific intensity is either taken from the atmosphere models computed by \citet{SPW12, Suleimanov.etal:17} for the given $\teff(\theta)$ and $g_{\rm eff}(\theta)$ or  approximated by a diluted blackbody 
given by Eq. (\ref{eq:emit_diluted_elsc}); 
the dilution factor $w$ and the colour-correction factor $\fc$ are found by interpolation in the existing grid of the precomputed values \citep[see][]{Suleimanov.etal:17}.  
The accuracy of this approximate method is demonstrated in Fig.\,\ref{fig:spectra} where the results  of the accurate computations taking real atmosphere models (solid curves) are compared with the approximate ones (dashed curves) for two relative luminosities ($\ell =$\,0.9 and 0.1), two inclination angles, $i=0\degr$ and 90\degr, and for two models of the flux distribution over the NS surface.
The errors in the bolometric fluxes are typically below 0.5\% and 1.5\% for the high- and low-luminosity models, respectively. 
They are not significant for the following discussion.

The final stage is the approximation of the observed spectrum from a rapidly rotating NS by a diluted blackbody $w'\uppi B_E(\fc' T_{\rm eff,\infty})$ in the observed energy range 3--20\,keV, where $T_{\rm eff,\infty}$ is the  redshifted effective temperature of the corresponding  non-rotating NS with the relative luminosity $\bar \ell$, which is an independent input parameter of the problem. 
The observed bolometric luminosity of the rotating NS model $L(\nu_*, i)$ can also be computed, together with the relative luminosity  $\ell'=L(\nu_*,i)/\ledd(0)$, where $\ledd(0)$ is the observed Eddington luminosity of the corresponding non-rotating NS of mass $M$. 
Repeating this procedure at a grid of $\bar \ell$  we obtain the model dependences $w' - \ell'$ and $\fc' - \ell'$, which are shown in Figs.\,\ref{fig:fcw_relative} and \ref{fig:fcw_absolute}. 
The dependence $w' - \ell'$ has to be compared with an observed dependence $K - F_{\rm BB}$ for some appropriate X-ray bursting NS. 
We chose the mass and the radius of a non-rotating NS to provide the surface gravity $\log g=$\,14.15.
Then we took the rotation frequency to be equal 700\,Hz, and computed the curves  $w' - \ell'$ and $\fc' - \ell'$ for three inclination angles, 0\degr, 45\degr, and 90\degr\ assuming solar chemical compositions with heavy element abundances reduced 100 times. The curves were computed for both limiting cases of the flux distribution over the surface, CRF  (Fig.\,\ref{fig:fcw_relative}) and CAF (Fig.\,\ref{fig:fcw_absolute}).
In the second case, the maximum  luminosities that can be reached correspond to the local Eddington limit at the NS equator.

The qualitative behaviour of the colour-correction factor $\fc'$ and the dilution factor $w'$ is similar to those shown in Fig.\,\ref{fig:corrections}. 
The colour correction shows a greater increase
for highly inclined NSs ($i=90\degr$), whereas the greater increase in the dilution factor occurs for the face-on NS ($i=0\degr$). 
The results for the CAF case and the  relatively low luminosity ($L/\ledd\sim 0.1$) are  very close quantitatively to the results shown in Fig.\,\ref{fig:corrections} ($R=14$\,km, the electron scattering case). 
In this case, the local diluted blackbodies at different latitudes are very close to each other. 
At the higher relative luminosities the local blackbodies differ more as the relative flux $F(\theta)/F_{\rm Edd}(\theta)$ near the equator approaches the local Eddington limit.

\subsection{Influence of NS rotation on determination of basic parameters}
\label{sec:qual_ns_rot} 

\begin{figure}
\centering{\includegraphics[width=0.82\columnwidth]{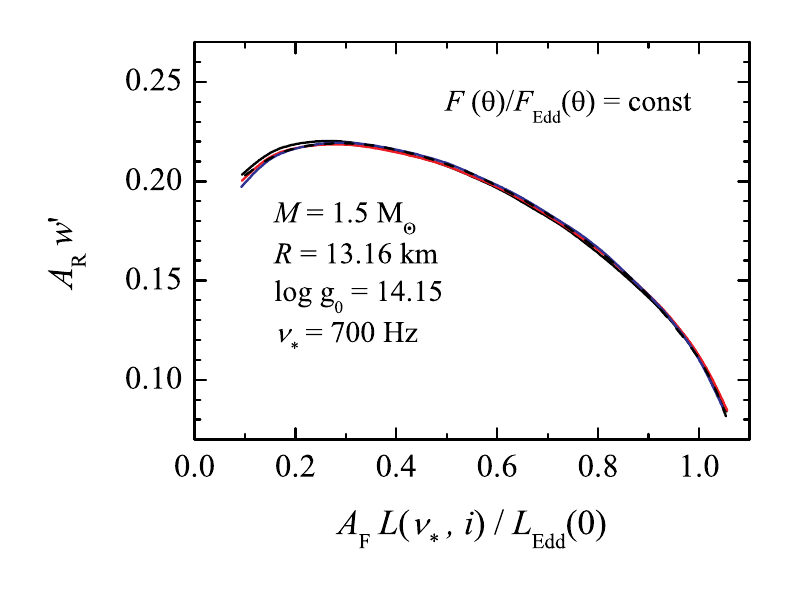}
\hspace*{0.3cm} \includegraphics[width=.8\columnwidth]{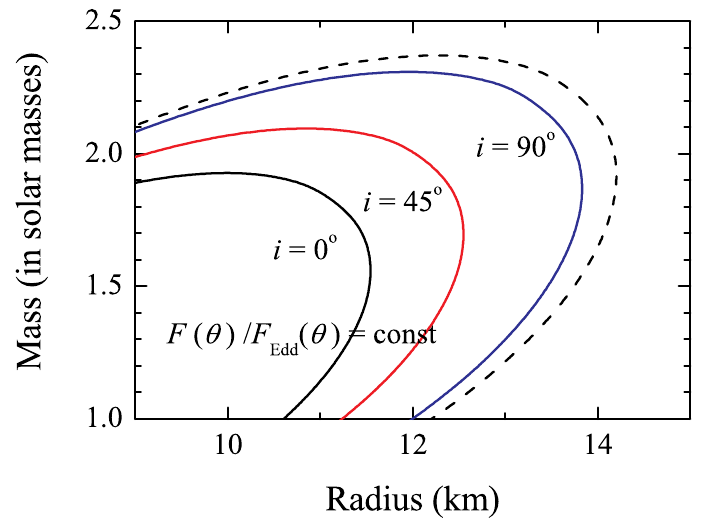}}
 \caption{\textit{Top panel}: Results of fitting the $w - \ell$ dependence with $w' - \ell'$ dependences computed for different inclination angles $i$ and assuming constant relative flux over the NS surface (CRF case, see Fig.\,\ref{fig:fcw_relative}). 
The corresponding correction factors are $(A_{\rm F},A_{\rm R})= (1.053,0.856)$  for $i=0\degr$, $(1.1,0.972)$ for 45\degr, and $(1.15,1.12)$ for 90\degr. 
\textit{Bottom panel:} Corresponding solutions at the $(M,R)$ plane obtained assuming a non-rotating NS  (dashed curve) and using additional correction factors $A_{\rm F}$ and $A_{\rm R}$ for different inclination angles.  }
\label{fig:MRcorr_relative}
\end{figure}

\begin{figure}
\centering{\includegraphics[width=0.82\columnwidth]{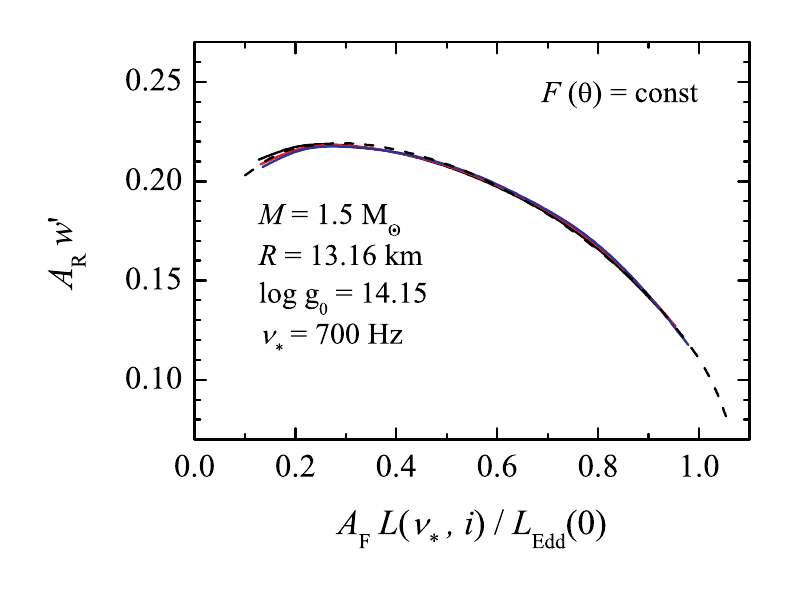}
\hspace*{0.3cm} \includegraphics[width=.8\columnwidth]{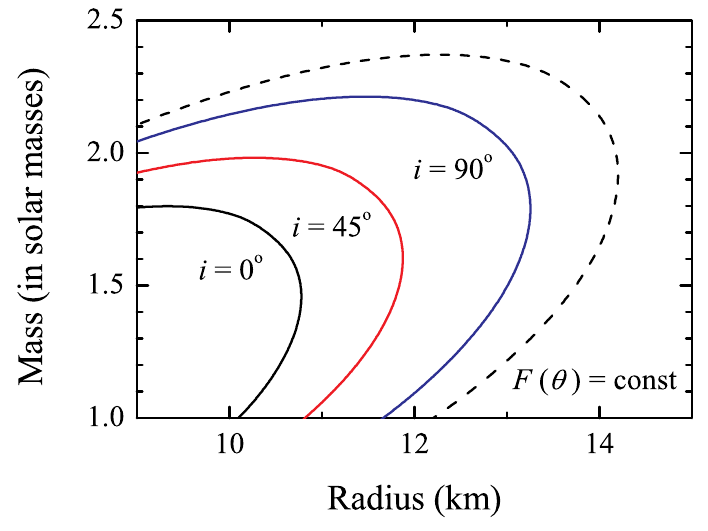}}
 \caption{Same as Fig.\,\ref{fig:MRcorr_relative}, but for constant absolute flux over the NS surface (CAF case, see Fig.\,\ref{fig:fcw_absolute}). 
The corresponding correction factors are $(A_{\rm F},A_{\rm R})= (1.12,0.85)$ for $i=0\degr$, $(1.16,0.97)$ for 45\degr, and $(1.2,1.12)$  for 90\degr. 
}
\label{fig:MRcorr_absolute}
\end{figure}

Using computed dependences $w' - \ell'$ we can estimate the effect of rapid rotation on determination of the NS radius.
Let us assume that the computed dependence $w - \ell$ for a non-rotating NS (see the dashed curves in the bottom panels of Figs.\,\ref{fig:fcw_relative} and \ref{fig:fcw_absolute}) closely fits the observed dependence $K - F_{\rm BB}$ for some X-ray burst occurring at the surface of a rapidly rotating NS with $\nu_* = 700$\,Hz. 
Let us also suggest that the fitting parameters (see Sect.\,\ref{sec:directcooling}) are $F_{\rm Edd,\infty} =$\,6.76$\times 10^{-8}$\,erg\,s$^{-1}$\,cm$^{-2}$ and $\Omega =$\,1261\,(km/10\,kpc)$^2$.
They can be combined to obtain a distance-independent quantity, the observed Eddington temperature, which corresponds to the dashed curve at the $M-R$ plane in the bottom panels of Figs.\,\ref{fig:MRcorr_relative} and \ref{fig:MRcorr_absolute}. 

It is possible to fit the dependence $w - \ell$ with the dependences $w' - \ell'$ computed for the rapidly rotating NS (see upper panels in Figs.\,\ref{fig:MRcorr_relative} and \ref{fig:MRcorr_absolute}) and to obtain the correction factors $A_{\rm F}$ and $A_{\rm R}$ to the fitting parameters $F_{\rm Edd,\infty}$ and $\Omega$. 
This means that the new parameters, which should actually be used to determine $M$ and $R$ of the non-rotating NS, are  $F'_{\rm Edd,\infty} = A_{\rm F} F_{\rm Edd,\infty}$ and $\Omega' = A_{\rm R} \Omega$. 
The new parameter $F'_{\rm Edd,\infty}$ is larger than the old one ($A_{\rm F} > 1$) for all inclination angles because the critical Eddington limit is reached at a rotating NS  at a lower luminosity. 
The new observed solid angle $\Omega'$ maybe smaller or larger than $\Omega$ depending on the inclination: 
$A_{\rm R} > 1$ for large inclination angles and  $A_{\rm R} < 1$ for a face-on rotating NS (see Figs.\,\ref{fig:fcw_relative} and \ref{fig:fcw_absolute}, bottom panels). 
The value of $A_{\rm R}$ depends on two factors. 
It depends first on the apparent area, which is always larger for a rotating NS (see Fig.\,\ref{fig:area}), which would make $A_{\rm R}$ smaller than unity.
In addition, the dilution factor depends on the colour temperature of the spectrum of a rotating NS.  
The higher the ratio of the colour to the local effective temperatures, the smaller the dilution factor $w'$. 
For highly inclined rapidly rotating NSs, the colour temperature is higher due to the Doppler boosting (see Fig.\,\ref{fig:corrections}), and its influence is more important than the increase in the apparent area resulting in $A_{\rm R} > 1$.

As a result the solutions at the $M-R$ plane for the different inclination angles are shifted (see Figs.\,\ref{fig:MRcorr_relative} and \ref{fig:MRcorr_absolute}, bottom panels).  
The interpretation of these shifts is as follows. If we use the non-rotating NS model  curve $w - \ell$, we obtain an incorrect solution presented by the dashed curve instead of the correct one shown by the solid curves for different  inclination angles. 
For example, if we treat a face-on rapidly rotating (with $\nu_* = 700$\,Hz) NS as a non-rotating NS, we   obtain that the NS radius $R$ is   3--3.5\,km larger than that obtained using the method that accounts for rotation.
For an edge-on system, the radius is nearly the same within 0.5--1\,km. 
We note here once more that the equatorial radius of the rapidly rotating NS is larger than the radius of the non-rotating NS with the same baryonic mass. 
For instance, a non-rotating NS with $R = 12$\,km and  $M = 1.5 M_\odot$, will have the equatorial radius $R_{\rm e} \approx 13$\,km if it rotates at $\nu_* = 700$\,Hz. 
Because there is an observational bias which restricts the inclination for real observed NSs in LMXBs to be $i < 70\degr$, treating NSs of unknown inclination as non-rotating will always result in an overestimate of the radius. 
We finally note that the two models for the flux distribution over the NS surfaces give  similar radius values, within 0.5\,km (compare Figs.\,\ref{fig:MRcorr_relative} and \ref{fig:MRcorr_absolute}), but the  CAF model  systematically gives a smaller radius.

\subsection{Essentials of the method}
\label{sec:essentials} 

Now we describe the method for obtaining NS parameters from the data, using the models for rotating NSs. 
We need to have a reasonable estimate of the rotation frequency $\nu_*$, the inclination, and the chemical composition of the atmosphere. 
Let us choose a pair of mass and radius  $(M,R)$ of the corresponding non-rotating  NS, and compute the model curve $w' - \ell'$ using the method described above. 
Then we fit the observed $K - F_{\rm BB}$ dependence with this model curve using  the same fitting parameters $F_{\rm Edd,\infty}$ and  $\Omega$ as for a non-rotating star. 
The corrections for the rapid rotation are accounted for by differences between the model curve $w' - \ell'$ and the corresponding model curve $w-\ell$ for a non-rotating NS. 
Both fitting parameters depend on the distance to the source $D$, which is actually the only fitting parameter for the given pair $(M, R)$. 

We repeat the fitting procedure at a grid $(M,R)$ obtaining the $\chi^2$ map which determines the most probable values of $M$ and $R$ of a non-rotating NS and corresponding confidence intervals. 
This map thus depends on the assumed values of the inclination angle $i$, the rotational frequency $\nu_*$, and   the chemical composition of the atmosphere.

\section{Application of modified cooling tail method to X-ray bursts from rapidly rotating NSs}
\label{sec:application}

\begin{figure*}
\centering{\includegraphics[width=0.8\columnwidth]{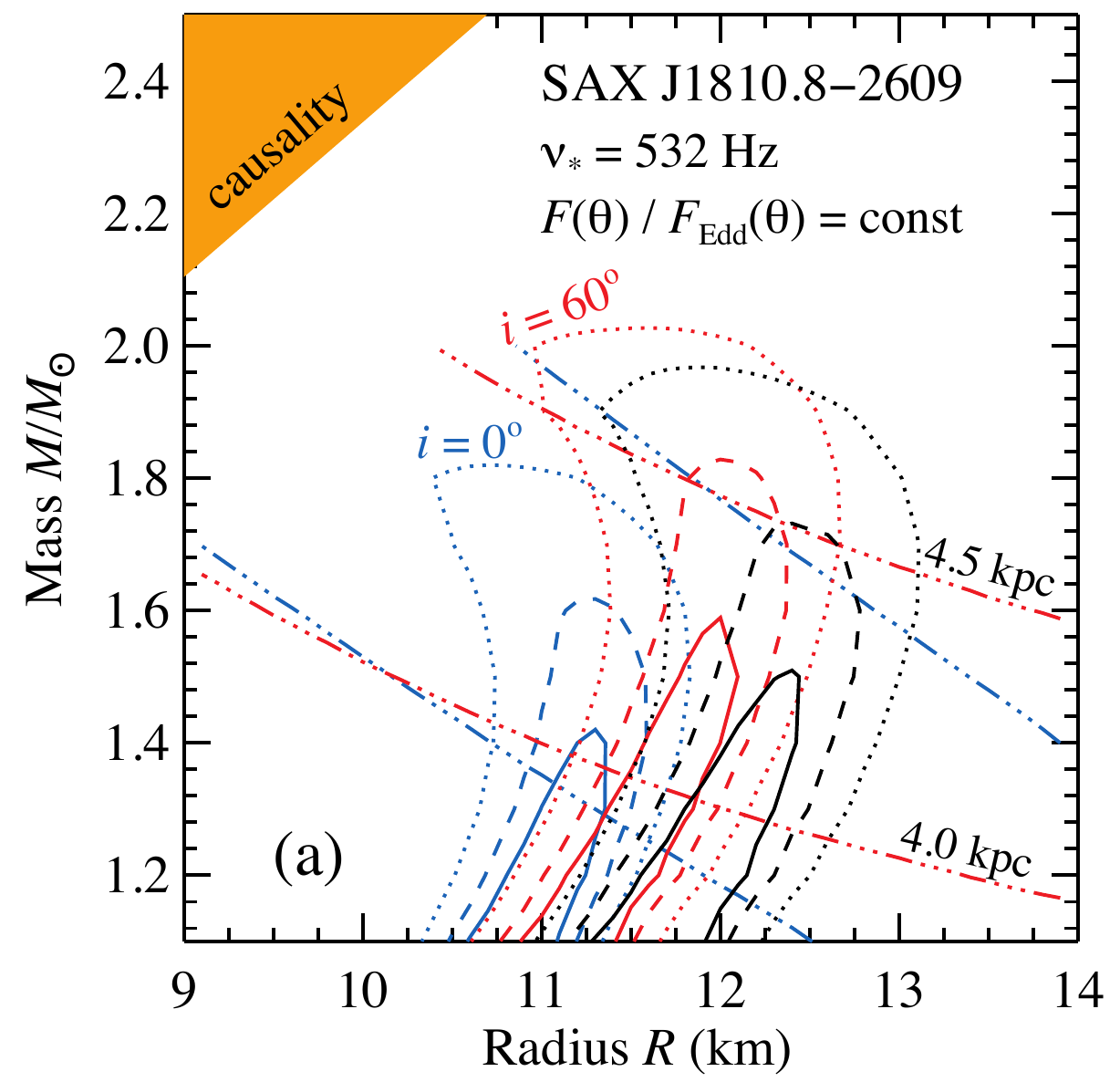}\hspace{1cm}
\includegraphics[width=.8\columnwidth]{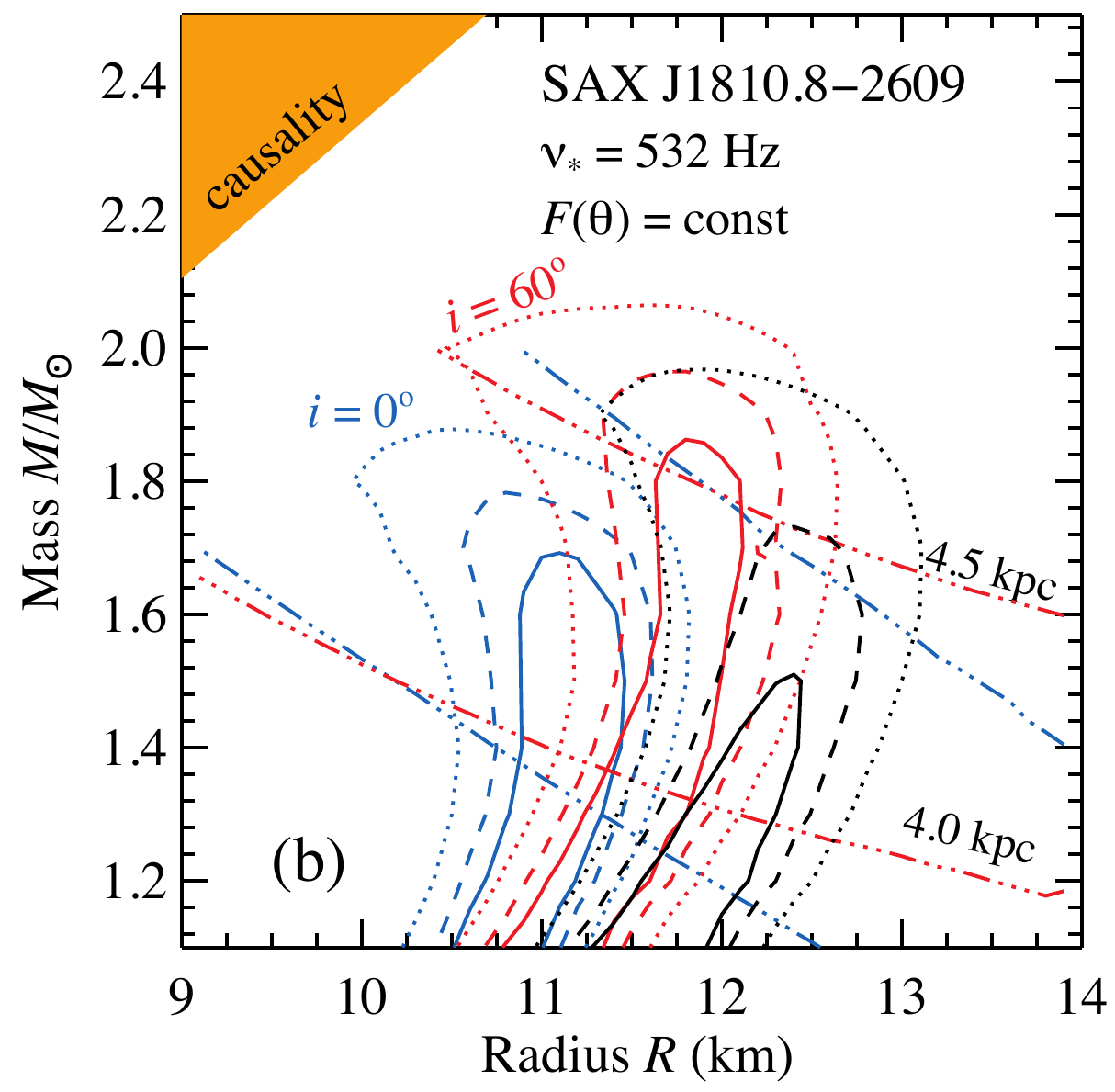}}
\caption{Confidence regions at the $M$--$R$ plane for SAX J1810.8$-$2609 obtained from the $\chi^2$ map using the cooling tail method modified for rapid rotation. 
The solid, dashed, and dotted curves indicate the 68\%, 90\%, and 99\% regions, respectively.
The red contours correspond to the inclination  $i$=60\degr, while the blue contours are  for the face-on NS. 
The black contours give the constraints assuming a non-rotating NS.
 (a) CRF case, i.e. constant relative flux. 
 (b) CAF case, i.e. constant absolute flux.
 The triple dot-dashed curves are the contours of the constant distance of 4.0 and 4.5\,kpc 
 (distance is the only free parameter for a given pair of $M$ and $R$). }
\label{fig:chi2}
\end{figure*}

SAX J1810.8$-$2609 was used as a key source for NS radius determination in two works \citep{Nattila.etal:16, Suleimanov.etal:17}, where we found the radius to be in the range 11.0--12.5 km assuming $M =1.5\msun$ and that the NS rotates slowly.
Unfortunately, there is only one X-ray burst that could be used for the cooling tail method. 
Recently, \citet{Bilous.etal:18} using archival \textit{RXTE} observations discovered  in that source burst oscillation at frequency $\nu_* =$\,531.8\,Hz.  
Such a high rotation rate may affect the radius determination significantly, and  we can use it as a test bed for our new modified cooling tail method.

The burst that we  use for our study has Obs ID 93044-02-04-00, and it was observed on 2007 August 13. 
The details about the burst, the data reduction, and blackbody fitting can be found in \citet{Nattila.etal:16}. 
The results of the application of the modified cooling tail method are shown in Fig.\,\ref{fig:chi2}. 
We consider three different cases. 
First of all, in order to test the new method and compare it with the previous results, we assumed that the NS is non-rotating, $\nu_* =$\,0\,Hz, but used the new method modified for rapid rotation. 
The resulting $\chi^2$ map, presented in Fig.\,\ref{fig:chi2} with solid black curves, is identical to the map produced with the direct cooling tail method for a non-rotating NS shown in Fig.\,5 in \citet{Suleimanov.etal:17}.
  
Then we computed $\chi^2$ maps taking the actually observed rotational frequency and assuming two inclination angles, $i =$\,0\degr\ and 60\degr. 
We considered two limiting cases for the flux distribution over the NS surface as described in Sect.\,\ref{sec:app_spectra}: the constant relative flux (CRF, Fig.\,\ref{fig:chi2}, left panel) and the constant absolute flux (CAF, Fig.\,\ref{fig:chi2}, right panel). 
In both cases the assumption about the face-on inclination gives the NS radius that is  approximately 1~km smaller than that assuming a non-rotating NS. 
The best estimate of the NS radius is 11.2$\pm$0.5\,km at $M =1.5\msun$,
and the corresponding equatorial radius $R_{\rm e}$ is 11.6$\pm 0.6$\,km. 
Taking a higher inclination of 60\degr\ leads to the $\chi^2$ maps closer to the map assuming no rotation and the NS radius of 11.8$\pm$0.5\,km at $M =1.5\msun$.
In this case the corresponding equatorial radius, 12.3$\pm 0.6$\,km, is close to the radius of the NS determined under assumption that it is non-rotating. 
We note that the CRF case gives smaller $\chi^2$ values than the CAF model, and also a smaller $\chi^2$ is achieved at higher inclination ($i = 60\degr$). 
However, the statistical significance of the improvement is below 3$\sigma$.

We recall that the only free parameter in the direct cooling tail method (in addition to the mass and radius) is the distance, which therefore can be estimated. 
The solutions that correspond to the distances of 4.0 and 4.5\,kpc are shown by the triple dot-dashed  lines in Fig.\,\ref{fig:chi2}. 
There is only one estimation for the distance to SAX J1810.8$-$2609 \citep{Natalucci.etal:00}, but it cannot be considered  as an independent one, as they used the maximum fluxes of type I X-ray bursts assuming that they correspond to the Eddington flux for a NS with $M = 1.4 M_\odot$. 
Nevertheless, their estimation of $4.9 \pm 0.3$\,kpc is close to ours.

Another X-ray bursting source where rapid rotation should be accounted for is 4U\,1608$-$52, which rotates at $\nu_* =$\,620\,Hz \citep{Muno.etal:02}.  
The cooling tail method applied to this source assuming no rotation gave the NS radius in the range 13--16\,km depending on the selected data points for $M =1.5\msun$ \citep{Poutanen.etal:14}.  
We found that the rapid rotation with $\nu_* =$\,700\,Hz can increase the visible NS radius by 2--3\,km for face-on systems and by 1\,km for edge-on systems.  
Thus, accounting for the rotation, the NS radius in 4U\,1608$-$52 will be in range 10--13\,km for low inclination and 12--15\,km for an inclination around $70\degr$. 
This is consistent with the current best limits on the NS radii 10.5--13.5\,km coming from other sources and observables \citep{Steiner13,Nattila.etal:16,Nattila.etal:17,Sul.etal:17,Abbott.etal:18}.

\section {Summary}
\label{sec:summary} 

We have presented a new version of the cooling tail method,  modified for rapid rotation, for determining NS parameters from X-ray bursts. 
The aim of the method is to find the most probable values for the mass and radius of a non-rotating NS, whose rotating model describes the observed spectral evolution of X-ray bursts taking place on the surface of a rapidly rotating NS.
The rotation frequency  $\nu_*$, the inclination angle of the NS rotation axis to the line of sight $i$, and the atmospheric chemical composition are the input parameters of the model. 

First, we   developed a model that transforms a non-rotating NS with given $M$ and $R$ to the approximate model of a rapidly rotating NS with the same baryonic mass of the increased equatorial radius $R_{\rm e}$ and the slightly increased gravitational mass $M'$. 
We   derived the approximate relations between $R$ and $R_{\rm e}$, and between $M$ and $M'$ using the models of rapidly rotating NSs computed by \citet{CShT94} for various EoS. 

At the next step we computed the emergent spectra of the obtained rotating NS models for different relative luminosities $\ell$ of the non-rotating NS with the given $M$ and $R$.   
As the effective temperature distribution over the rotating NS surface at any burst moment is not known a priori, we considered two limiting cases: CRF (i.e. the same ratio $\ell$ of the bolometric flux to the local Eddington flux) and CAF (i.e.  the same absolute bolometric flux at each latitude, the same as for the non-rotating NS with a given $\ell$). 
We note that the local Eddington flux depends on the effective gravity which includes the centrifugal force.
We showed that approximating NS atmosphere spectra by a diluted blackbody gives a good description of the observed spectra from a rotating NS.
Therefore, for further modelling we used a diluted blackbody approximation for the  local spectra. 
The parameters of these spectra, the dilution factor $w$ and the colour-correction factor $\fc$, are found by interpolating in the precomputed set of models at a grid of the effective temperatures, effective surface gravities, and  chemical compositions of the atmosphere from  \citet{SPW12, Suleimanov.etal:17}. 

We used the universal (independent of EoS) slow-rotation approximations for description of the shape  of the rotating NS and the effective surface gravity distribution suggested by \citet{mors07} and \citet{ALGM:14}. 
To compute the observed spectra and integrate over the visible surface of a rapidly rotating NS, we used a simplified approach computing approximately the total redshift factor using a metric of the rotating NS in a slow-rotation approximation and making corrections for the oblate shape \citep[see][]{mors07, NP:18,SNP18}. 
The light bending angle and the lensing factor were computed using accurate analytical formulae from \citet{Poutanen19}, which allowed us to speed up calculations by orders of magnitude. 

The computed spectra are then fitted with the diluted blackbody model. 
The fitting parameters, the dilution factor $w'$, and the colour-correction factor $\fc'$, are found from comparison with the fiducial spectrum, which was the blackbody spectrum with the effective temperature of a non-rotating NS with the same relative luminosity $\ell$. 
We thus obtained the model dependence $w' - L(\nu_*,i)/\ledd(0)$, where $\ledd(0)$ is the Eddington luminosity of a non-rotating NS. 
This model curve  can  then be compared to the observed dependence $K -F_{\rm BB}$ using the distance to the source as the only fitting parameter, in addition to  $M$ and $R$ of a non-rotating NS.
As a result we get a $\chi^2$-map at the $M-R$ plane, which allows us to determine the most probable values and the confidence region for $M$ and $R$ of the corresponding non-rotating NSs.
These constraints depend on the assumed inclination $i$,  rotation rate $\nu_*$, and chemical composition. 

We applied our method to an X-ray burst  observed from SAX~J1810.8$-$2609 and investigated earlier by \citet{Nattila.etal:16} and \citet{Suleimanov.etal:17}, who assumed no NS rotation. 
We revised the NS radius determination using our modified cooling tail method accounting for rapid rotation of this NS with $\nu_* =$\,531.8\,Hz  \citep{Bilous.etal:18}. 
We found that the NS radius is smaller by $\approx$1\,km (11.2$\pm$0.5\,km instead of 12.2$\pm$0.5\,km assuming no rotation) for the face-on inclination and $M=1.5\msun$. 
The corresponding equatorial radius $R_{\rm e}$ of this rapidly rotating NS, 11.6$\pm 0.6$\,km,  is still slightly smaller than the radius of the NS obtained assuming no rotation.
At high inclination the effect of rapid rotation is small.

\section*{Acknowledgments} 

This research has been supported by the grant 14.W03.31.0021 of the Ministry of Science and Higher Education of the Russian Federation.  
V.F.S. also thanks Deutsche  Forschungsgemeinschaft  (DFG) for financial support (grant WE 1312/51-1).  
We thank the German Academic Exchange Service  (DAAD, project 57525212), the Academy of Finland (projects 317552 and 331951) and the Magnus Ehrnrooth foundation for travel grants. 

\bibliographystyle{aa}
\bibliography{references}

\appendix

\section{Shape and gravity of a rotating neutron star}
\label{sec:app1}

Let us consider a rotating NS with the gravitational mass $M'$, equatorial radius $R_{\rm e}$, and observed angular velocity $\Omega_*=2\uppi \nu_*$. 
For completeness, we present here the whole collection of formulae from \citet{ALGM:14} that we use to define the NS shape and gravity.
They demonstrate that basic properties of the rotating NS model can be represented using two dimensionless parameters
\be
       x=\frac{GM'}{c^2\, R_{\rm e}},
      \qquad \bar{\Omega} = \Omega_*\,\left(\frac{R_{\rm e}^3}{GM'}\right)^{1/2}.
\ee
The metric of a stationary axisymmetric rotating NS is usually written in the form \citep{BI:76}  
\bea
\label{eq:metr}
 ds^2&= &-e^{2\nu}\, c^2dt^2 +\bar{r}^2\,\sin^2\theta\,B^2\,e^{-2\nu}(d\phi-\varpi\,dt)^2
\nonumber \\ &+&e^{2\zeta-2\nu}(d\bar{r}^2+\bar{r}^2\,d\theta^2) ,
\eea 
where metric coefficients depend on coordinates $\bar{r}$  and co-latitude $\theta$. 
The circumferential radius is related to $\bar{r}$ by \citep{FIP:86}
\be\label{eq:rBenur}
       r = B\, e^{-\nu}\, \bar{r}.
\ee
The metric coefficients for a non-rotating stationary NS are known as the isotropic Schwarzschild metric coefficients 
\bea
\label{eq:metr2}
\nu_0 &=& \ln\frac{1-\frac{{\bar u}}{2}}{1+\frac{{\bar u}}{2}}, \nonumber \\ 
B_0& = & \left(1-\frac{{\bar u}}{2}\right)  \left(1+\frac{{\bar u}}{2}\right), \\  
\zeta_0 & =& \ln B_0 , \nonumber 
\eea
where 
\be \label{eq:ubar_rbar}
{\bar u}= \frac{GM'}{c^2\bar{r}} . 
\ee
A well-known relation exists between the circumferential $r$ and isotropic Schwarzschild radial coordinate $\bar{r}$: 
\be
       r = B_0 e^{-\nu_0}\, \bar{r} = \bar{r}\,\left(1+\frac{{\bar u}}{2}\right)^2 .
\ee 
The metric coefficients in Eq.\,(\ref{eq:metr}) can be expanded in powers of $\bar{\Omega}$ up to order 2  \citep{ALGM:14}
\bea
\label{eq:metr1}
\nu &=& \nu_0 + \left(\frac{b}{3} -q\,P_2(\cos\theta)\right) {\bar u}^3 , \nonumber \\ 
B& = & B_0 +b {\bar u}^2, \\  
\zeta & =& \zeta_0 + b\,\left(\frac{4}{3}P_2(\cos\theta)-\frac{1}{3}\right)\, {\bar u} ^2, \nonumber 
\eea
where $P_2$ is the Legendre polynomial of order 2, and $b$ and $q$ are the dimensionless coefficients
\bea
   b &=&  0.4454\,\bar{\Omega}^2\, x, \\ 
   \label{eq:quadr}
   q &=&  -0.11\,\frac{\bar{\Omega}^2}{x^2}. 
\eea
For a given NS described by parameters $x$ and $\bar{\Omega}$,  for given circumferential radius $r$ and angle $\theta$, 
the metric coefficients are computed using Eqs.\,(\ref{eq:rBenur}), (\ref{eq:metr2}), (\ref{eq:metr1}), and (\ref{eq:ubar_rbar})  via iterations.
The angular velocity of the local zero angular momentum observer with respect to an observer at rest at infinity $\varpi$ can  also be expanded with the same accuracy 
\be \label{eq:angvel_ZAMO}
\varpi =\frac{2G \cal J}{c^2\,\bar{r}^3}\,(1-3 {\bar u}), 
\ee
where ${\cal J} = I \Omega_*$ is the NS angular momentum, $I(M',\Omega_*)=\bar{I} \,M' R_{\rm e}^2$ is the moment of inertia and 
\be
   \bar{I} =  x^{1/2}\,(1.136-2.53\,x + 5.6\,x^2). 
\ee
Eq. (\ref{eq:angvel_ZAMO}) can be written in dimensionless form as
\be \label{eq:angvel_ZAMO_dim}
\bar{\varpi} = \varpi \,\left(\frac{R_{\rm e}^3}{GM'}\right)^{1/2} 
= 2 \frac{{\bar u}^3 (1-3 {\bar u})}{x^2}\ \bar{I}\ \bar{\Omega}.
\ee

 \begin{figure}
\centering
{\includegraphics[width=0.8\columnwidth]{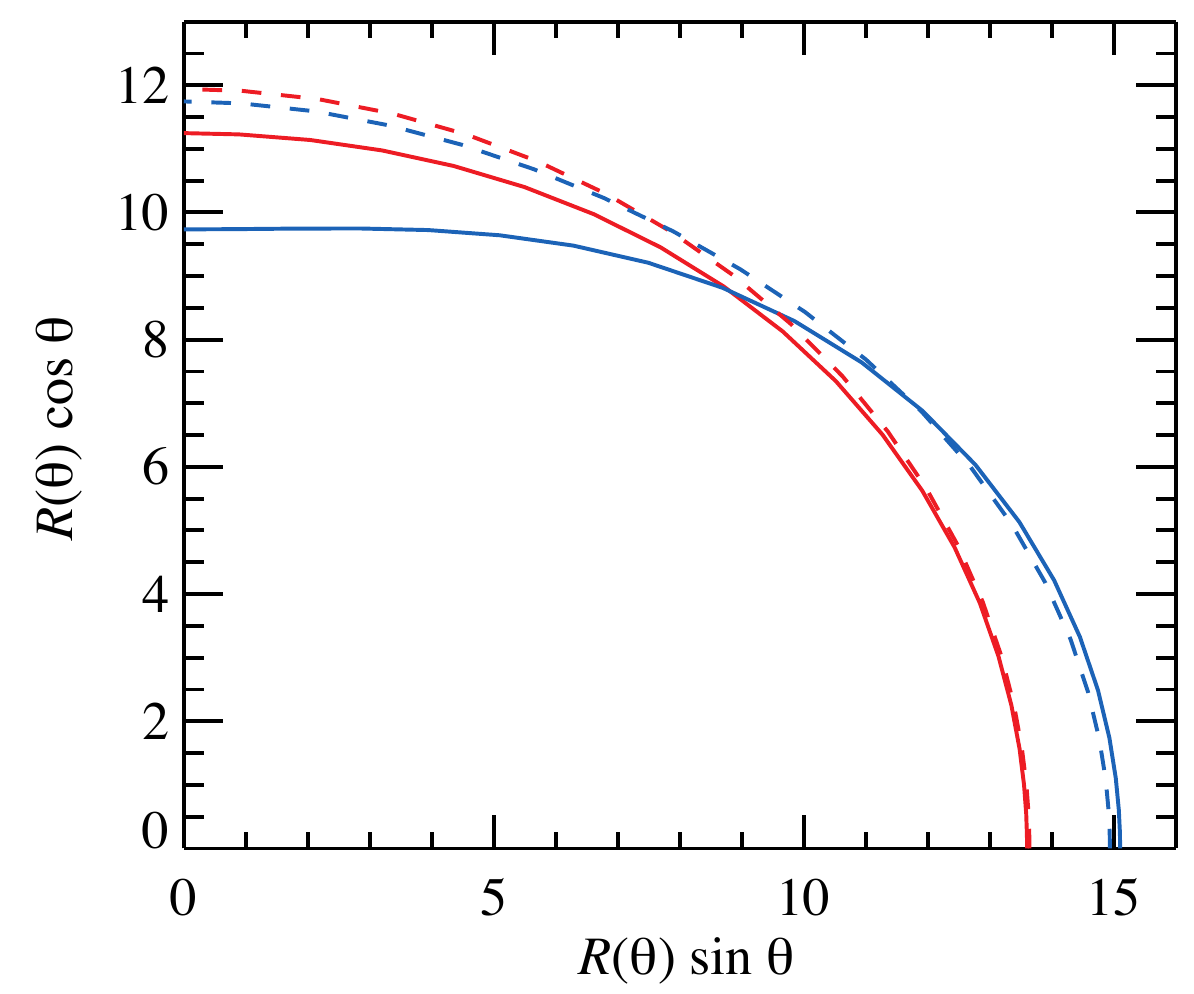}

\vspace*{0.5cm}
\includegraphics[width=0.8\columnwidth]{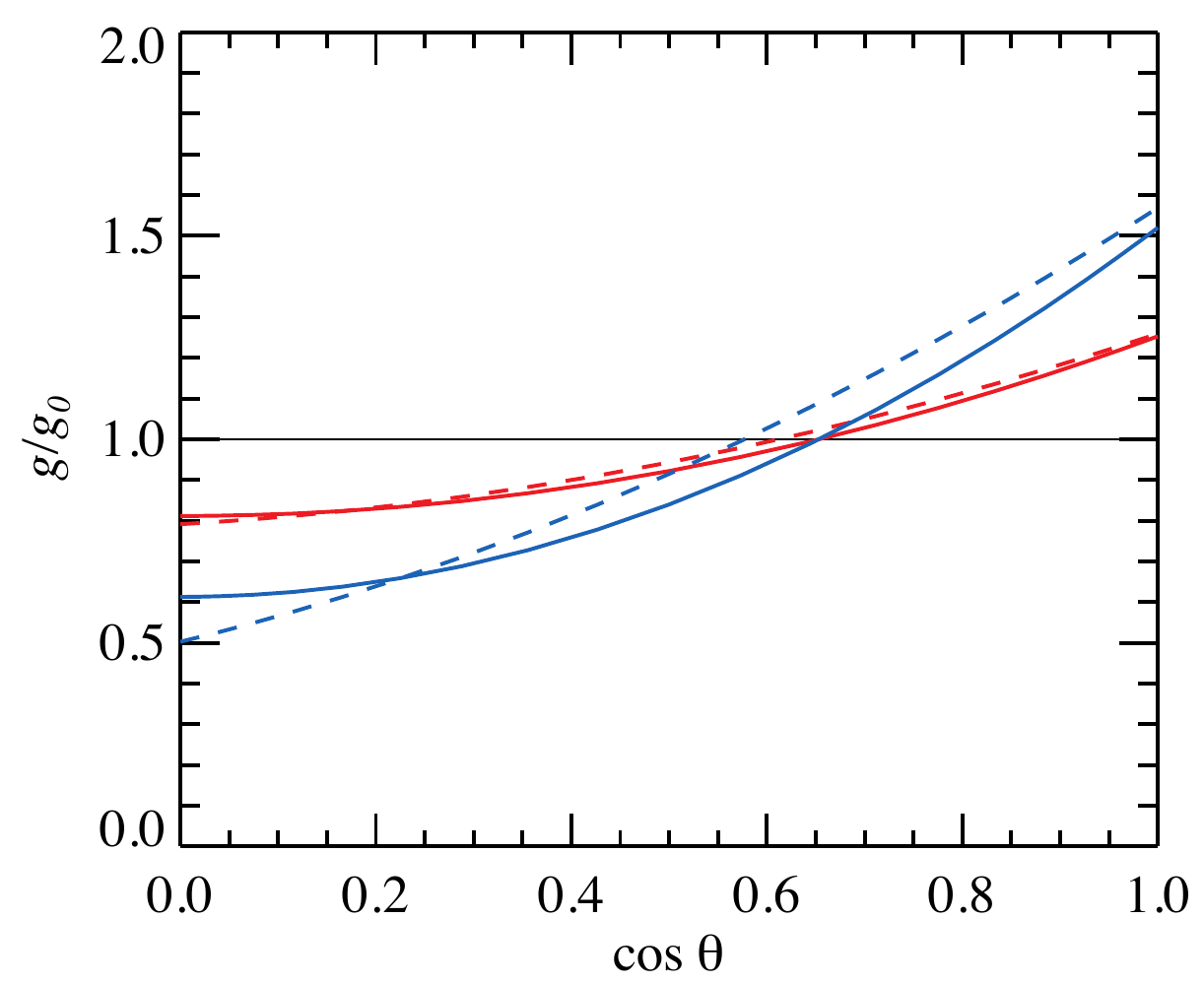}}
 \caption{
\textit{Top panel}: Shape of the rotating NS computed with two different approximations.
The solid curves correspond to Eq.\,(\ref{eq:rteta}), while the dashed curves to Eqs.\,(8)--(10) in \citet{mors07}.
\textit{Bottom panel}: Effective surface gravity distributions. 
The solid curves correspond to Eq.\,(\ref{eq:grav}), while the dashed curves to Eq. (50) in \citet{ALGM:14}. 
The red curves correspond to a relatively slow rotating NS with $\bar \Omega = 0.3$ ($\nu_*=842$\,Hz, $M'=1.8\msun$, $R_{\rm e} = 13.6$\,km).
The blue curves are for a faster rotating NS with  $\bar \Omega = 0.6$ ($\nu_*=1084$\,Hz, $M'=1.99\msun$, $R_{\rm e} = 15.1$\,km).
}
 \label{fig_shape}
\end{figure}

Following \citet{ALGM:14}  we use   the approximation for the circumferential radius of a rapidly rotating NS  
\be \label{eq:rteta}
   R(\theta) =R_{\rm e} \,\left[1 -\bar{\Omega}^2(0.788-1.03\,x)\, \cos^2\theta\right], 
\ee
and for the effective gravity in a slow-rotation approximation 
\be \label{eq:grav}
g(\theta) = g_0\left(1+c_{\rm e}\bar \Omega^2\,\sin^2\theta +c_{\rm p}\bar \Omega^2\,\cos^2\theta\right),
\ee
where
\be
     g_0 = \frac{GM'}{R_{\rm e}^2}\,\left(1-2x\right)^{-1/2}
\ee
is the surface gravity for a spherical NS with mass $M'$ and radius $R_{\rm e}$, and 
\be \label{eq:grav2}
   c_{\rm e} = 0.776\,x-0.791,\qquad c_{\rm p} = 1.138-1.431\,x.
\ee   

We note that the approximate fitting formulae for the effective surface gravity (\ref{eq:grav}) and the rapidly rotating NS shape (\ref{eq:rteta}) are not self-consistent as they are independent approximations to the accurately computed values. 
Slightly different approximations for the  gravity and for the shape are given by Eq. (50) in \citet{ALGM:14} and Eqs. (8)--(10) in \citet{mors07}, respectively. 
A comparison of these approximations to those given by Eqs. (\ref{eq:grav}) and (\ref{eq:rteta}) for two rotating NSs is shown in Fig.\,\ref{fig_shape}. 
The approximation to the NS shape derived by \citet{mors07} (dashed curves) gives a less oblate NS shape.
The difference in the effective gravity approximations is insignificant for slowly rotating NSs. 
We also compared the shape with the models published by \citet{CShT94}.
For example,  for $M' = 1.4103\msun$, $R_{\rm e} = 11.71$\,km, and $\Omega = 5033.6$\,s$^{-1}$ ($\nu_*= 801.12$\,Hz), \citet[][see their Table 15]{CShT94} give the eccentricity $\epsilon = 0.488$, while Eq.\,(\ref{eq:rteta}) gives  $\epsilon = 0.495$, which is about 2\% accurate.  
We conclude that we can use approximations  (\ref{eq:rteta}) and (\ref{eq:grav}) for slowly rotating NSs with $\bar \Omega < 0.3$.

\section{Radiation observed  from a rotating neutron star}
\label{sec:app2}

Accurate computations of the observed radiation from a rapidly rotating NS are rather complicated. 
The main reason is a complex form of geodesics in the vicinity of the star, which requires using the full ray tracing  approach \citep[see e.g.][]{Baubock.etal:12, NP:18}.
This is a time-consuming approach that is not appropriate for the extensive computations presented in this work. 
A simplified approach was proposed by \citet{mors07}. 
They considered a local  Schwarzschild metric at every surface point of an oblate NS. 
They computed gravitational redshift and the light bending effect as for a non-rotating NS adding the Doppler boosting factor that accounts for rotation  (oblate Schwarzschild or OS approach). 
This approach gives acceptable results for slowly rotating NSs with $\nu_* < 700$\,Hz.  
However, we believe that it is possible to account for the main effects introduced by the metric of the rapidly rotating NS without significant complication of the OS approach cited above. 
\citet{NP:18} showed that it is important to have an accurate treatment of the redshift effect (i.e. the ratio of the observed to the emitted photon energy  $E/E'$), which is affected both by the non-zero quadrupole momentum and by the frame dragging impact on the Doppler boosting factor. 
Thus, we introduce here a new modified OS (MOS) approximation, using the metric of a rapidly rotating NS instead of the local Schwarzschild metric for the photon energy change, but keeping simplified Schwarzschild description for the light bending, which is the main  simplification comparing to the exact treatment. 

Let us start with discussing the emission of a rapidly rotating spherical NS in Schwarzschild metric. 
Then we consider the correction coming from the shape of the NS. 
Finally, the correction to the redshift due to the  non-zero quadrupole momentum and by the frame dragging is  discussed. 
 
\begin{figure}
\centering
\includegraphics[width=1.\columnwidth,trim = {2.5cm 1.5cm 1.cm 1cm}, clip]{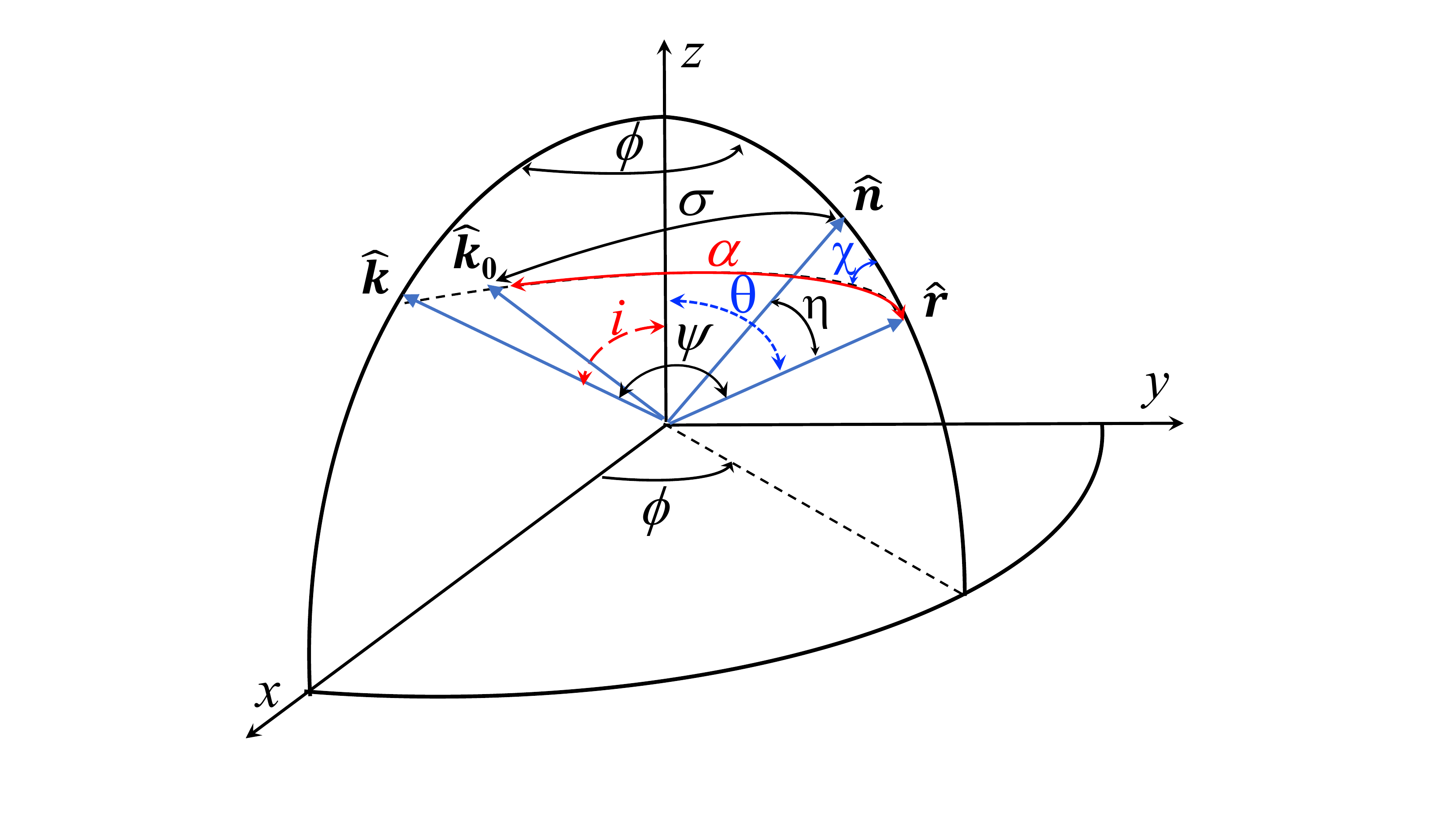} 
\caption{Geometry of the problem.}
\label{fig:geom}
\end{figure}

\subsection{Rapidly rotating spherical star} 
\label{sec:app2_sph}

The flux observed by an observer can be computed in two different ways. 
We can integrate over the spherical surface in the non-rotating frame, or make integration over the surface in the co-rotating frame of the NS. 
Both approaches should  produce the same result. 

Let us start from the first approach. 
We introduce a spherical coordinate system with the polar axis along the NS rotation axis. 
An observer is situated at distance $D$ in the direction given by the unit vector $\unit{k}=(\sin i, 0, \cos i)$, with $i$ being the inclination (see Fig.\,\ref{fig:geom}).  
A surface element in the static frame can be defined in terms of co-latitude $\theta$ and azimuthal angle $\phi$ with the unit vector of its position being $\unit{r}=(\sin\theta\cos\phi, \sin\theta\sin\phi, \cos\theta)$. 
Its area is $dS=R^2 d\cos\theta\, d\phi$. 
The solid angle of this element as observed at distance $D$ is 
\be\label{eq:omegaobs}
d\Omega_{\rm obs} = \frac {\bar b d\bar b\,d\varphi}{D^2} ,
\ee 
where the impact parameter $\bar b$ and the polar angle $\varphi$ are defined using polar coordinates on the image plane centred to the observed NS centre. 
The impact parameter is related to the emission angle $\alpha$, the angle between the radial direction $\unit{r}$, and the unit vector along photon momentum $\unit{k}_0$ close to the NS surface as \citep[see e.g.][]{Beloborodov:02}
\be \label{eq:impact_alpha}
\bar b = \frac{R}{\sqrt{1-u}} \sin\alpha , 
\ee
where $u=R_{\rm S}/R$ is the NS compactness with $R_{\rm S}=2GM/c^2$ being the Schwarzschild radius. 
Substituting Eq.\,(\ref{eq:impact_alpha}) to Eq.\,(\ref{eq:omegaobs}) and given that the element area can   also be written as $dS=R^2 d\cos\psi\, d\varphi$, with $\psi$ being  the angle between the local radius-vector $\unit{r}$ at the NS surface and the line-of-sight $\unit{k}$  in flat space-time, 
\be 
\cos\psi= \unit{r} \cdot \unit{k}=  \cos\theta  \cos i +  \sin\theta\sin i\,\cos\phi ,
\ee
we get \citep{Beloborodov:02,PB06} 
\be \label{eq:dOmegaobs}
d\Omega_{\rm obs} =  {\cal D} \cos \alpha \frac{dS}{D^2} 
\ee  
with the lensing factor
\be
{\cal D} =\frac{1}{1-u}\frac{d \cos \alpha}{d \cos \psi} .
\ee
The relation between $\alpha$ and $\psi$ is given by the integral \citep{mtw73,PFC83,Beloborodov:02}
\be \label{eq:bend}
  \psi=\int_R^{\infty} \frac{dr}{r^2} \left[ \frac{1}{\bar b^2} -
       \frac{1}{r^2}\left( 1- \frac{R_{\rm S}}{r}\right)\right]^{-1/2} . 
\ee
It can be computed using an accurate numerical procedure (e.g. as is given in the Appendix of \citealt{SNP18}), but here  we use for simplicity an approximate formula suggested in \citet{Poutanen19}
\be\label{eq:poutanen19app}
\cos\alpha \approx 1 - y \,(1-u)\,  G(y,u),
\ee
where $y=1-\cos\psi$, and 
\be \label{eq:Gfunc}
    G(y,u) = 1+\frac{u^2\,y^2}{112}-\frac{e}{100}u\,y\,\left[\ln\left(1-\frac{y}{2}\right)+\frac{y}{2}\right].
\ee
This formula is an improvement over the analytical approximation suggested by \citet{Beloborodov:02} and is 0.05\% accurate over the whole range of angles of interest and for any realistic NS compactness.
Using this  relation we obtain the expression for the lensing factor
\be \label{eq:calD}
{\cal D} =1+\frac{3u^2 \,y^2}{112} 
-\frac{e}{100}u \,y\,\left[2\ln\left(1-\frac{y}{2}\right)+y\frac{1-3y/4}{1-y/2}\right].
\ee 
The unit vector along photon momentum close to NS surface can be found from 
\be 
\unit{k}_0= \frac{\sin\alpha\ \unit{k}+ \sin(\psi-\alpha)\ \unit{r}}{\sin\psi} . 
\ee 

The observed flux at energy $E$ from the element will be 
\be \label{eq:dFE}
dF_E= I_E  d\Omega_{\rm obs} .
\ee
The observed specific intensity can be related to the intensity measured in the co-rotating frame as 
\be\label{eq:IE_EE}
I_{E} = \left (\frac{E}{E'}\right )^3 I'_{E '} (\alpha',\theta) ,
\ee
where the ratio of energies combines gravitational redshift and the Doppler effect
\be \label{eq:EEpr_sph}
 \frac{E}{E'}  = \delta\, e^{\nu_0}  =  \delta \sqrt{1-u},
\ee
and in principle we can consider the possibility that the local intensity also depends on co-latitude $\theta$ and the zenith angle $\alpha'$ measured in the corotating frame of the element. 
The Doppler factor 
\be \label{eq:Doppler_sph}
\delta = \frac{1}{\gamma(1-\beta\cos\xi)} 
\ee
depends on the NS velocity at this latitude relative to the external non-rotating frame,
\be \label{eq:beta_only}
\beta(\theta) = \frac{R\Omega_*}{c \sqrt{1-u}} \sin\theta ,
\ee 
and the Lorentz factor is 
\be  \label{eq:Lorentzgamma}
\gamma(\theta)=\frac{1}{\sqrt{1-\beta^2(\theta)}}. 
\ee
The angle $\xi$ between the photon momentum and the spot velocity vector $\vec{\beta}=\beta (-\sin\phi,\cos\phi,0)$ in the external static frame can be expressed as \citep{PG03,PB06} 
\be
\cos\xi = \unit{\beta}\cdot \unit{k}_0 = - \frac{\sin\alpha}{\sin\psi} \sin i \,\sin\phi.
\ee
The angle $\alpha'$ in Eq.\,(\ref{eq:IE_EE})  the photon momentum makes to the radial direction as measured in the co-rotating frame of the spot  is related to the similar angle measured in the static frame as 
\be
\cos\alpha'=\delta\cos\alpha .
\ee 
We finally obtain the total observed flux by integrating Eq.\,(\ref{eq:dFE}) over the whole visible stellar surface 
\bea \label{eq:FluxTotal}
F_{E} &=&   \int_{\cos \alpha>0} \left(\frac{E}{E'}\right)^3I_{E'} (\alpha', \theta)\,d\Omega_{\rm obs}  \\
& = &\frac{R^2}{D^2}  \int\! \! d\cos\theta \! \!  \int\! \!   d\phi  \cos \alpha\ {\cal D}
\left(\delta\sqrt{1-u}\right)^3   I'_{E '} (\alpha',\theta) . \nonumber
\eea

Now let us consider an alternative derivation when we integrate over the NS surface in the co-rotating frame. 
Let us define an element on the NS surface at co-latitude $\theta$ and azimuthal angle $\phi'$ in the frame corotating with the NS, with the extent given by $d\theta$ and $d\phi'$. 
The area of this element is $dS'=\gamma R^2 d\cos\theta\,d\phi'$, where the Lorentz factor $\gamma$ appears because the area expressed in angular coordinates is measured by co-moving observers \citep{NP:18,Lo18,Bogdanov19L26}.
The solid angle $dS'$ occupies on the observer's sky is given by Eq.\,(\ref{eq:dOmegaobs}) with the only difference that $\cos\alpha\,dS$ should be substituted by $\cos\alpha'\,dS'$, so that we obtain 
\be \label{eq:dOmegaobs_prime}
d\Omega_{\rm obs} = {\cal D} \cos \alpha\  \frac{\delta\ \gamma\ R^2 d\cos\theta\,d\phi'}{D^2}. 
\ee  
The observed flux is thus  \citep{PB06}
\be \label{eq:dFE_obs}
dF_{E} (\phi_{\rm obs} ) =   
\frac{R^2}{D^2}  d\cos\theta \  d\phi'  \cos \alpha\ {\cal D}
\left(\delta \sqrt{1-u}\right)^3  \delta \ \gamma   I'_{E '} (\alpha',\theta) . 
\ee
We note here that it is important  to distinguish between the emission phase $\phi$ (i.e. position of the element when photons were emitted as measured by external static observer) and the arrival (observed) phase $\phi_{\rm obs}$ when the  flux  from this element is actually observed.  
Their differentials are  simply related by the standard time contraction formula as  \citep{RL79} 
\be \label{eq:timecontr}
d \phi_{\rm obs} = (1-\beta\cos\xi) d\phi . 
\ee
Now averaging the flux given by Eq.\,(\ref{eq:dFE_obs}) over the observed phase  and changing the integration variable to the emitted phase using  Eq.\,(\ref{eq:timecontr}) we obtain 
\bea
\langle dF_{E} \rangle & =  &  \frac{1}{2\uppi} \int_0^{2\uppi}dF_{E} (\phi_{\rm obs} ) \  d\phi_{\rm obs} \\ 
&=&   \frac{R^2}{D^2}  d\cos\theta\, \frac{d\phi'}{2\uppi}  \!\! \int_0^{2\uppi}\!\!\!\!\! d\phi \cos \alpha\ {\cal D} \ 
\left(\delta\sqrt{1-u}\right)^3     I'_{E '} (\alpha',\theta) . 
 \nonumber
\eea
We now notice that the factor $\gamma(1-\beta\cos\xi)$ has cancelled out with one of $\delta$. 
Further integration over $\phi'$  reduces to just removing the factor $1/{2\uppi}$  because the flux does not depend on the choice of $\phi'$. 
Thus, for the total flux observed from a star we arrive at Eq.\,(\ref{eq:FluxTotal}).

\subsection{Rapidly rotating oblate star} 
\label{sec:app2_obl}

For the oblate star with the radius depending on co-latitude as $R(\theta)$, we need to make small modifications to the formulae presented in the previous section \citep{mors07,SNP18}. 
Now the local normal $\unit{n}$ does not coincide with the radial direction, and the angle between them $\eta$ (see Fig.\,\ref{fig:geom}) can be expressed as 
\be
\cos\eta = \unit{n}\cdot \unit{r} = \frac{1}{\sqrt{1+f^2(\theta)}} , 
\ \sin\eta = \frac{f(\theta)}{\sqrt{1+f^2(\theta)}} , 
\ee
where 
\be
f(\theta) =  \frac{1}{\sqrt{1-u(\theta)}} \frac{1}{R(\theta)}\frac{dR(\theta)}{d\theta} , 
\ee
\be 
u(\theta)=\frac{2GM'}{c^2 R(\theta)},
\ee
and the derivative  of the circumferential radius could be found from Eq.\,(\ref{eq:rteta}): 
\be
\frac{dR(\theta)}{d\theta} = 2R_{\rm e}\bar{\Omega}^2(0.788-1.03x) \sin\theta\, \cos\theta.
\ee
The area of the surface element (as measured in the static frame) is 
\be 
dS= \frac{1}{\cos\eta} R^2(\theta)d\cos\theta\, d\phi ,
\ee
where the factor $\cos\eta$ accounts for deviation of the stellar surface from the spherical surface.

The local atmosphere of a rotating oblate NS is axisymmetric relative to the local normal. 
Therefore, we have to transform the angle $\alpha$ measured relative a local radius-vector to the angle $\sigma$ measured relative to a local normal. 
This angle can be found from the spherical law of cosines
\be\label{eq:cos_sigma1}
 \cos \sigma = \cos\eta \cos\alpha +  \sin\eta\sin\alpha\cos\ \chi,
\ee
where  
\be
\cos \chi = \frac{\cos i -\cos\theta\cos\psi }{\sin\theta \sin\psi}, 
\ee
and thus 
\begin{eqnarray}\label{eq:cos_sigma2}
 \cos \sigma & =&  \cos\eta \cos\alpha \nonumber \\
 &+&   \frac{\sin\alpha}{\sin\psi} \sin\eta\ (\cos i \sin\theta -\sin i\cos\theta\cos\phi ). 
\end{eqnarray}
Similarly to the spherical case correcting for relativistic aberration, we obtain the emission zenith angle in the frame comoving with the surface 
\be
 \cos\sigma' = \delta \cos\sigma. 
\ee
Thus, we arrive at a modified expression for the solid angle   
\be \label{eq:dOmegaobs_oblate}
d\Omega_{\rm obs} = {\cal D}\ \frac{\cos \sigma}{\cos\eta} \ \frac{R^2(\theta)\,d\cos\theta\, d\phi}{D^2}. 
\ee  
If we know the local specific emergent radiation intensity $I_{E'}(\sigma', \theta)$ at the surface of a rotating NS as a function of energy $E'$, co-latitude $\theta$, and the angle  $\sigma'$ between the emergent ray and the surface local normal, we can compute the total observed spectrum $F_{E}$ from the whole surface 
\be \label{eq:FE_oblate}
F_{E} =   \int_{\rm \cos \sigma>0} \left(\frac{E}{E'}\right)^3 \ I'_{E'} (\sigma', \theta)\,d\Omega_{\rm obs}.
\ee
We note here in the formulae for the light bending angle (\ref{eq:poutanen19app}) and the lensing factor (\ref{eq:calD}),  both $u$ and $R$ are now latitude-dependent.  
In the calculations presented in this paper we considered different cases for the local  specific intensity $I'_{E'} (\sigma', \theta)$ as described in Sect.\,\ref{sec:surface_spectra}. 
The total photon energy change now can be expressed as 
\be  \label{eq:EEpr_drag}
 \frac{E}{E'}  = \delta\, e^{\nu} (1+\beta'\cos\xi) . 
\ee
The three terms in this expression correspond to the Doppler factor $\delta$, 
a pure gravitational redshift $e^{\nu}$ (which accounts for quadrupole moment, see Eq. \ref{eq:metr1}), and a phenomenological factor with 
\be
\beta' = \frac{R(\theta)\varpi}{c}e^{-\nu}\sin\theta 
\ee
being the frame dragging dimensionless velocity. 
The expression for $\varpi$ is given by Eq. (\ref{eq:angvel_ZAMO}). 
In the formulae for the Lorentz (\ref{eq:Lorentzgamma}) and Doppler factors (\ref{eq:Doppler_sph}), instead of surface velocity given by Eq. (\ref{eq:beta_only})  we use the velocity relative to a zero angular momentum observer: 
\be
 \beta(\theta) = \frac{R(\theta)}{c}\,e^{-\nu}\,(\Omega_* -\varpi)\,\sin\theta.
\ee

These ad hoc formulae nevertheless give accurate values of the total energy redshift $1+z\equiv E'/E$ at the pole and for the maximum and minimum values at the equator for an observer at $i=90\degr$. 
For example, for a NS model published in Table 15 of \citet{CShT94} with $M' = 1.4103\msun$, $R_{\rm e} = 11.71$\,km, and $\Omega = 5033.6$\,s$^{-1}$ (i.e. $\nu_*= 801.12$\,Hz), our redshift at the pole $z_{\rm p} =0.2796$ is lower by just 0.1\% than the value $z_{\rm p} =0.280$ given there. 
Our maximum and minimum redshifts at the equator (for $i=90\degr$), $z_{\rm b} =0.6094$ and $z_{\rm f}= -0.0244$, are very close to those computed by \citet{CShT94}, $z_{\rm b} =0.606$ and $z_{\rm f}= -0.025$.

\begin{figure}
\centering
\includegraphics[width=0.8\columnwidth]{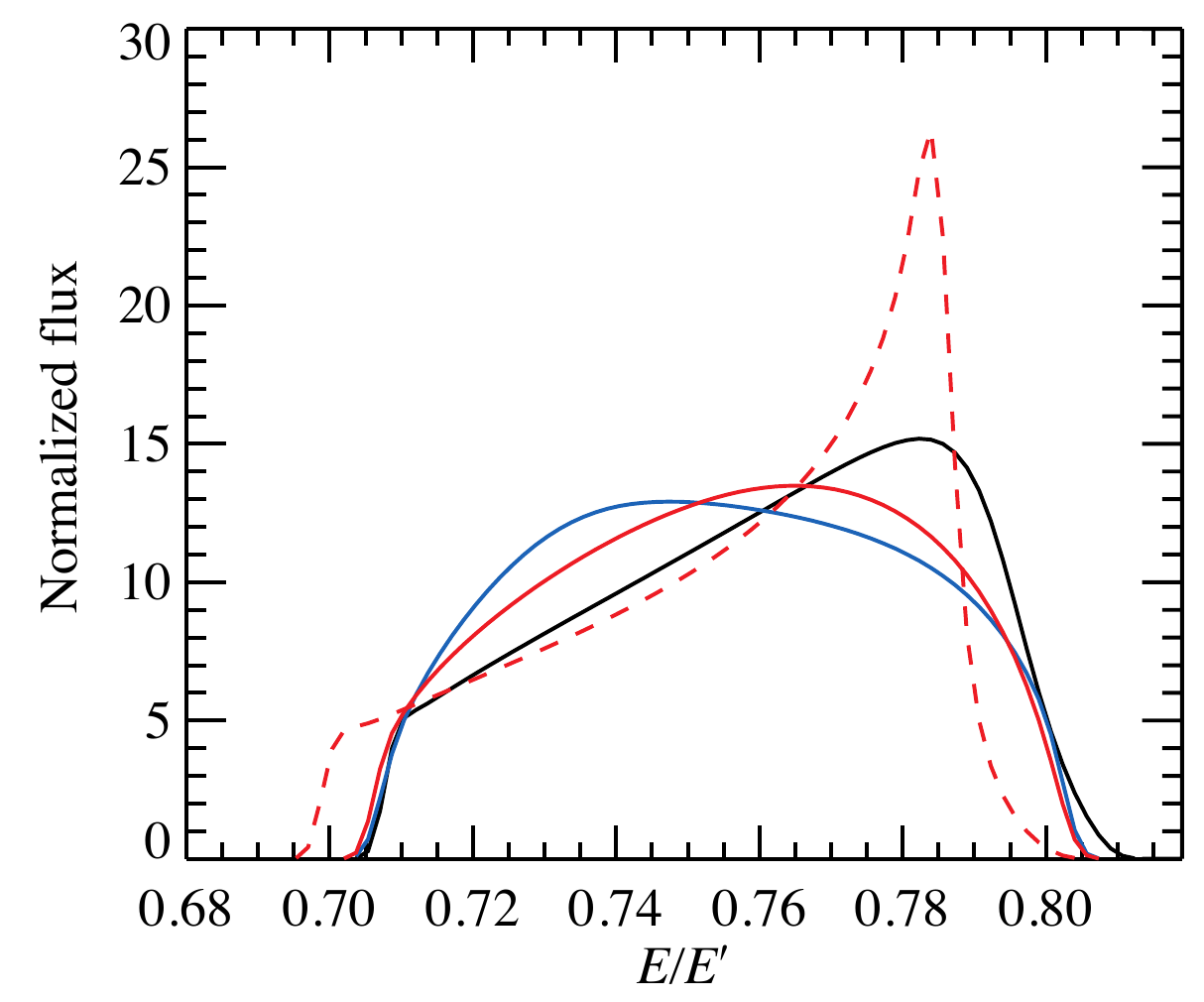} 
\caption{Line profiles from a NS with $M' = 1.4\msun$, $R_{\rm e} = 10$\,km rotating at 700 Hz, and seen by a distant observer at  inclination  $i = 20\degr$. 
The line profiles for a spherical NS with $R=R_{\rm e}$ and for an oblate NS shape both with  Schwarzschild exterior metric are shown by the black and blue curves, respectively. 
Accounting for frame dragging and quadrupole moment gives the profile shown by the red curve. 
The line obtained with a quadrupole moment artificially increased by a factor of 4 above that given by Eq.\,(\ref{eq:quadr}) is shown by the dashed red curve. 
Compare to fig.\,5 in \citet{NP:18}.  }
\label{fig:line_method}
\end{figure}

\begin{figure}
\centering
\includegraphics[width=0.8\columnwidth]{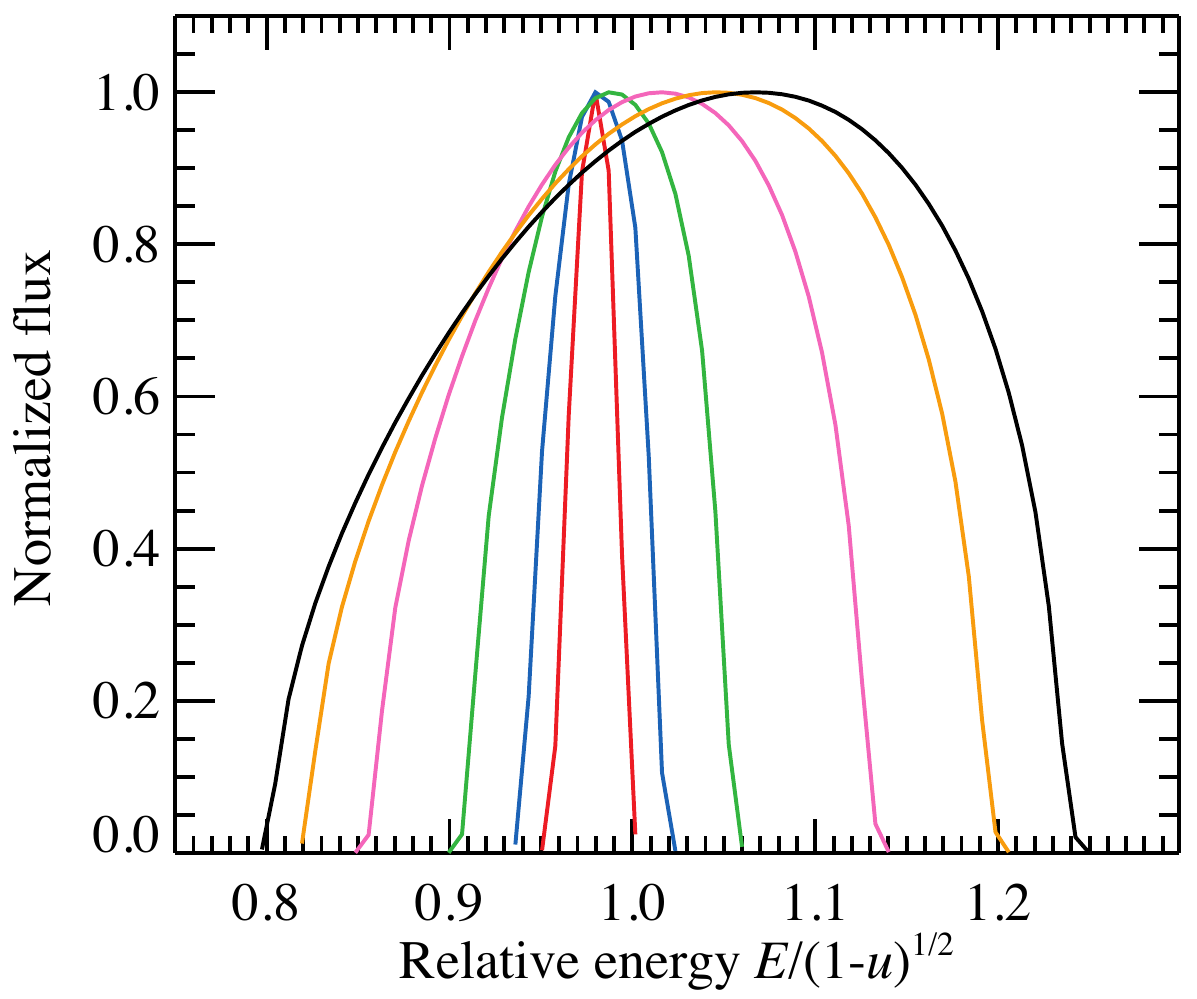} 
 \caption{Line profiles observed at different inclination angles, $i=90\degr$ (black), 60\degr\ (orange), 40\degr\ (violet), 20\degr\ (magenta), 10\degr\ (blue), and 5\degr\ (red). 
The adopted NS parameters are $M'=1.5\msun$, $R_{\rm e}= 14$\,km, $\nu_* = 600$\,Hz. 
The energy range is shifted by the redshift factor $\sqrt{1-R_{\rm S}/R_{\rm e}}$.
Compare to fig.\,6 in \citet{NP:18}.  }
\label{fig:line_incl}
\end{figure}

We also computed the observed profiles of a narrow emission line emitted by a rapidly rotating NS. 
In particular, we reproduced fig.\,5 from \citet{NP:18}, where  $M'=1.4\msun$, $R_{\rm e} = 10$\,km, $\nu_* = 700$\,Hz, and $i=20\degr$ were assumed. 
They computed the line profile using four different set-ups: the exact solution, the exact solution with the quadrupole moment artificially increased by four times, an oblate NS in local Schwarzschild metric, and  a spherical rotating NS ($R=R_{\rm e}$) in Schwarzschild metric with Doppler boosting taken into account. 
As intrinsic surface emission we took a Gaussian line  at unit energy with the standard deviation $\sigma=0.002$.  
Our results presented in Fig.\,\ref{fig:line_method} demonstrate that our approach allows us to reproduce the main features of the correct solution.  
The line profiles computed at different inclination angles are shown in Fig.\,\ref{fig:line_incl} for $M'=1.5\msun$, $R_{\rm e} = 14$\,km, and $\nu_*=600$\,Hz. 
The results closely reproduce  the main features of the  line profiles  shown in fig. 6 of \citet{NP:18}.

\label{lastpage}

\end{document}